\documentclass[referee]{raa}           

\usepackage{graphicx,times}
\usepackage{natbib}
\usepackage{amssymb,amsmath}

\bibpunct{(}{)}{;}{a}{}{,}
\newcommand{\hi}{{\rm H}{\textsc i}}

\usepackage[pagebackref=true]{hyperref}

\begin{document}

\title{ HI Vertical Structure of Nearby Edge-on Galaxies from CHANG-ES}

\volnopage{ {\bf 20XX} Vol.\ {\bf X} No. {\bf XX}, 000--000}
\setcounter{page}{1}

\author{Yun Zheng\inst{1}, Jing Wang\inst{1},  Judith Irwin\inst{2},  Q. Daniel Wang\inst{3}, Jiangtao Li\inst{6,7}, Jayanne English\inst{4}, Qingchuan Ma\inst{1},Ran Wang\inst{1}, Ke Wang\inst{1}, Marita Krause\inst{5}, Toky H. Randriamampandry\inst{1}, Rainer Beck\inst{5}
}
\institute{ Kavli Institute for Astronomy and Astrophysics, Peking University, Beijing 100871, China; {\it jwang$\_$astro@pku.edu.cn}\\
\and
Department of Physics, Engineering Physics, Astronomy, Queen's University, Kingston, ON, K7L 3N6, Canada\\
\and
Astronomy Department, University of Massachusetts, Amherst, MA 01003, USA\\
\and 
Department of Physics and Astronomy, University of Manitoba, Winnipeg, Manitoba, R3T 2N2, Canada\\
\and 
Max-Planck-Institut für Radioastronomie,Auf dem Hügel 69,53121 Bonn,Germany\\
\and
Department of Astronomy, University of Michigan, 311 West Hall, 1085 S. University Avenue, Ann Arbor, MI, 48109-1107, USA\\
\and
Purple Mountain Observatory, Chinese Academy of Sciences, 10 Yuanhua Road, Nanjing 210023, China\\
\vs \no
{\small Received 20XX Month Day; accepted 20XX Month Day}
}

\abstract{We study the vertical distribution of the highly inclined galaxies from the Continuum Halos in Nearby Galaxies – an EVLA Survey (CHANG-ES). We explore the feasibility of photometrically deriving the $\hi$ disk scale-heights from the moment-0 images of the relatively edge-on galaxies with inclination $>$80 deg, by quantifying the systematic broadening effects and thus deriving correction equations for direct measurements. The corrected $\hi$ disk scale-heights of the relatively edge-on galaxies from the CHANG-ES sample show trends consistent with the quasi-equilibrium model of the vertical structure of gas disks. The procedure provide a convenient way to derive the scale-heights and can easily be applied to statistical samples in the future.
}

\authorrunning{Y. Zheng et al.}            
\titlerunning{HI Vertical Structure of Nearby Edge-on Galaxies}  
\maketitle

\section{Introduction}
The vertical structure of $\hi$ gas discs is an important tracer of the galactic potential and dynamical effects in spiral galaxies.
Under the assumption of hydrostatic equilibrium, the $\hi$ gas vertical structure exhibits a force balance between the self-gravity of the disk and the effective pressure of $\hi$ \citep{boulares1990galactic, piontek2007models,  koyama2009pressure, ostriker2010regulation, krumholz2018unified}.
The effective pressure, largely supported by the turbulence \citep{mac1999energy, tamburro2009driving}, further reflects an energy balance between the radiative dissipation and the energy input that comes from stellar feedback and galactic-scale gas inflows driven by non-axisymmetric torques \citep{krumholz2010dynamics, forbes2012evolving, forbes2014balance}.
Real gas discs can be additionally influenced by gas accretion \citep{kerevs2005galaxies, dekel2006galaxy, oosterloo2007cold, dekel2009cold, zheng2017hst}, and tidal interactions \citep{1972ApJ...178..623T, 1992ARA&A..30..705B, 2007A&A...468...61D}, as well as energetic feedback from active nuclei \citep{2007MNRAS.380..877S, 2012ARA&A..50..455F, 2014A&A...562A..21C}.  

Observationally, the vertical structure of an $\hi$ disk is quantified by the scale height to first order \citep{2021ApJ...916...26R}, which reflects properties consistent with the hydrostatic equilibrium model.
The $\hi$ scale heights are nearly constant in the inner region of the Galaxy and increase appreciably as a function of radius (the so-called flaring phenomenon) in the outer region between 5 and 35 kpc \citep{dickey1990hi,kalberla2009hi}.
\cite{narayan2002vertical} explained this radial trend of the Galactic $\hi$ scale heights and emphasized the importance of gravitational coupling between gas, stars, and dark matter.
They concluded that in addition to the contributions from stars the gravity from the atomic and molecular gas helps set the scale heights in the inner region, while the atomic gas is more important than the molecular gas at intermediate radius; at large Galactic radius, the $\hi$ scale heights flare because the gravitational force decreases quickly and the dark matter halo dominates the gravity.
The sharpness of the $\hi$ flares at large galactic radius is explained by the truncation of the stellar disks \citep{van1988three}.
$\hi$ flares are found to be common in spiral galaxies  \citep{brinks1984high, bigiel2012universal}, though they are not as sharp as in the Milky Way, as truncated stellar disks are not observed in all galaxies \citep{bland2005ngc}.

The vertical hydrostatic equilibrium model of the $\hi$ disk serves as a useful tool to derive properties involved in this equilibrium.
Typically, for face-on galaxies, observational studies easily obtain the gas velocity dispersion but rely on the equilibrium model when deriving the gas disk thickness, while for edge-on galaxies, the situation is the other way round.
\cite{olling1995usage} and \cite{narayan2005constraints} developed a model using the flare of $\hi$ to constrain the shape, mass, and size of the dark matter halo.
\cite{2018MNRAS.477.2716K} concluded the isothermal nature of gas based on the vertical force balance between the gravitational drag force of gas, stars, and dark matter and the gas thermal, turbulent, and magnetic pressure. 
Recent developments include that \cite{2019A&A...622A..64B} derived the $\hi$ volumetric density to study the volumetric star formation law.
They found that the volumetric star formation law is much tighter than the surface star formation law, particularly, the previously known break in the slope of the surface star formation law \citep{bigiel2008star, leroy2008star} is more likely due to disc flaring rather than a decrease of the star-forming efficiency at low surface densities \citep{2019A&A...622A..64B}.

The vertical structure of the $\hi$ disk is more complex than described by this simple hydrostatic equilibrium model.
A thick $\hi$ layer, also called extraplanar gas or $\hi$ halo in the literature, with a typically 1$-$2 kpc scale height (in contrast to 100$-$200 pc for the thin disk) is directly observed in edge-on galaxies, with a lag in rotation with respect to the thin $\hi$ disc \citep{2007AJ....134.1019O, 2013MNRAS.434.2069K}.
The Galactic intermediate-velocity clouds can be viewed as a form of extraplanar gas \citep{oort1970formation, wakker1997high}. 
The extraplanar gas is also modeled and found to be prevalent in external galaxies that are not edge-on \citep{2001ApJ...562L..47F}. 
\cite{marasco2019halogas} did a systematic study of extraplanar gas in 15 nearby late-type galaxies and concluded that both the mass and kinematics of extraplanar gas are in good agreement with the galactic fountain model which is powered by stellar feedback \citep{shapiro1976consequences, bregman1980galactic}.
In the fountain model, gas is pushed away from the disk by the supernova feedback and falls back with extra gas from the circumgalactic medium (CGM) after metal enrichment.

More complexities come from the fact that the edge-on views of $\hi$ disks are not necessarily flat but warped outside the edge of optical disks \citep{1957AJ.....62...90B, 1957AJ.....62...93K, 1976A&A....53..159S, 1977MNRAS.181..573N, 1978PhDT.......195B, 1990ApJ...352...15B}.
Warps are found to be ubiquitous in disk galaxies \citep{1976A&A....53..159S, garcia2002neutral}.
The fact that they usually onset at the edge of optical disks suggests that the inner flat disk and the outer warped disk have different formation histories and probably involving different epochs \citep{van2007truncations}. 
The mainstream explanation for the formation of warps seems to be the accretion of material with an angular momentum vector misaligned with that of the main disk \citep{jiang1999warps, shen2006galactic, rovskar2010misaligned}. 
Other explanations include the torques from the misaligned inner disc and the associated inner oblate halo \citep{toomre1983theories, dekel1983galactic, sparke1988model}, and the tidal force from companion galaxies \citep{weinberg1995production, weinberg2006magellanic}.

An observational census would usefully gain more insights into the physics that shape the vertical extent of $\hi$ disks.
We use the highly inclined galaxies from the Continuum Halos in Nearby Galaxies – an EVLA Survey (CHANG-ES) \citep{2012AJ....144...43I, wiegert2015chang, 2019AJ....158...21I}. The $\hi$ information of 19 edge-on galaxies in CHANG-ES is published in \cite{2022MNRAS.513.1329Z}. Two galaxies are for the first time presented in $\hi$ interferometric images and twelve of galaxies have better $\hi$ spatial resolutions and/or sensitivities of intensity maps than literature. The $\hi$ data is not well resolved kinematically, but we manage to derive the radially averaged scale height for the 15 most edge-on galaxies (inclination $> 80^{\circ}$, section 3) with a similar method as another CHANG-ES study \cite{2018AA...611A..72K} used to measure the scale height of continuum halos. In the literature, the scale-heights have been derived with sophisticated kinematical modellings based on data of much higher spectral resolutions, particularly when the galaxies are less inclined \citep{2014AJ....148..127Y, 2020MNRAS.494.4558Y}. Most of those modeling methods need to assume a quasi-equilibrium between the gravity and the gas pressure, and thus are most accurate for unperturbed galaxies. The procedure to derive the scale-heights in this paper is mostly photometric using images, but we carefully correct the measurements for several types of observational artifacts, also called systematic biases or systematic broadening effects, including PSF smearing, planar projection, and edge-on projection. (section 3.2 and 3.3). Thus the procedure and measurements presented in this paper provide an alternative and convenient way to derive the scale-heights without significant assumptions for the dynamic states of gas. It can easily be applied to statistical samples in the future, and may potentially provide an indicator for the dynamic states of $\hi$ gas when compare to results or expectations from kinematically derived scale-heights. We investigate possible dependence of the $\hi$ scale height on other galactic properties (section 4), and also more closely discuss the uncertainties and effects of the external environment (section 5). We assume a $\Lambda$CDM Cosmology (H$_{0}$=73 km s$^{-1}$ Mpc$^{-1}$) and Kroupa initial mass function \citep{kroupa2001variation}.

\section{Data}
\subsection{Sample and HI data}
Continuum Halos in Nearby Galaxies – an EVLA Survey (CHANG-ES) is a deep radio continuum survey at 1.5 GHz (L band) and 6 GHz (C band), targeting 35 nearby edge-on galaxies \citep{2012AJ....144...43I}. 
The galaxies were observed with the The Karl G. Jansky Very Large Array (VLA) using the B, C, and D configurations in the C and L bands, and the observational details have been previously described by \cite{2012AJ....144...43I} and \cite{wiegert2015chang}.
CHANG-ES paper XXV \citep[][Z21 hereafter]{2022MNRAS.513.1329Z} produced the $\hi$ data at L-band C-configuration observation through the program of Common Astronomy Software Applications (CASA) \citep{2007ASPC..376..127M}. 19 galaxies in CHANG-ES sample were successfully reduced into $\hi$ data cubes, which have an average beam size of $\sim14.5''$ in full width half maximum (FWHM), typical velocity resolution of 52.8 km/s, and an average RMS of $\sim$0.4 mJy/beam. This sample is dominated by star-forming, $\hi$-rich galaxies. 

The resolution and depth of $\hi$ data are not sufficient for us to distinct thin $\hi$ disks from the thick $\hi$ disks, so we only investigate the averaged scale heights of the whole $\hi$ disks.
We study the $\hi$ scale heights in galaxies with inclinations larger than $80^{\circ}$, based on the same the criteria adopted by \citet{2018AA...611A..72K} who investigated the scale height of continuum halos from the CHANGE-ES sample. This criterion is to mitigate the contamination from projected disk plane. A total of 15 galaxies in CHANG-ES $\hi$ sample meet this criterion. The basic information of the 15 galaxies is listed in Table \ref{table1:basic}. 
We take the coordinates, the optical size ($R_{25}$, the 25 mag arcsec$^{-2}$ isophote semi-major axis in the $B$ band) from CHANG-ES Paper I \citep{2012AJ....144...43I}, and distances from CHANG-ES Paper IV \citep{wiegert2015chang}. The $\hi$ information, including $\hi$ mass ($M_{\rm HI}$), $\hi$ radius ($R_{\rm HI}$), position angle (PA) of $\hi$ disk, FWHM and RMS of $\hi$ intensity maps, are obtained from Z21. The $\hi$ radius was calculated through the $\hi$ size-mass relation from \cite{2016MNRAS.460.2143W}.

\subsection{SFR and mass densities}

We take the star formation rate (SFR) from \cite{2019ApJ...881...26V}, who calculated SFR based on fluxes from narrow-band H$\alpha$ images and fluxes from the Wide-field Infrared Survey Explorer 22 $\mu$m images. Because NGC 5084 lacks H$\alpha$ observation, we exclude this galaxy from SFR related analysis. The SFR values are listed in Table \ref{table3:hHI}, in which the SFR of NGC 5084 is only estimated by 22 $\mu$m data.
The star formation surface density is estimated as $\Sigma_{\rm SFR} = {\rm SFR}/ (\pi* R_{25}^2)$.

The total (or dynamic) mass surface density within the optical radius $\Sigma_{\rm Mtot,r25}$ is taken from Paper IX \citep{2018AA...611A..72K}, who calculated it based on the inclination corrected line widths of the $\hi$ spectrum.

We also derive the baryonic mass surface density within $R_{25}$, $\Sigma_{\rm Mbaryon, r25}$. We take the stellar mass $M_*$ from Z21, who derived it based on the i-band luminosity and g-i color correlated mass-to-light ratio \cite{2003ApJS..149..289B}. 
We further measure the $\hi$ mass within $R_{25}$, $M_{\rm HI,r25}$. 
The baryonic mass is calculated as the sum of the stellar mass and gas mass within $R_{25}$, $M_{\rm baryon,r25} = M_{*,r25} + 1.4 M_{\rm HI,r25}$, where 1.4 is the standard correction factor to account for helium and metals. Then the baryonic mass surface density is calculated as $\Sigma_{\rm baryon} = M_{\rm baryon,r25}/ (\pi* R_{25}^2)$. These surface densities are listed in Table \ref{table3:hHI}.

\begin{table}
\bc
\begin{minipage}[]{100mm}
\caption[]{Galaxy Sample\label{table1:basic}}\end{minipage}
\setlength{\tabcolsep}{1pt}
\small
 \begin{tabular}{ccclccccrcccccc}
  \hline\noalign{\smallskip}
Galaxies & R.A. (J2000)$^{a}$ & Decl. (J2000)$^{a}$ & $i^{b}$ & Distance$^c$ & $R_{25}^d$ & log$M_{\rm HI}$$^e$ & $R_{\rm HI}$$^e$ & PA$^f$ & FWHM$^g$ & RMS of image$^g$\\
 & [h m s] & [$^{\circ}$ $'$ $''$] & [{$^{\circ}$}] & [Mpc] & [kpc] & [$M_{\odot}$] & [kpc] & [{$^{\circ}$}] & [$''$] & [$10^{20} \rm cm^{-2}$]\\
  \hline\noalign{\smallskip}
NGC 2683 & 08 52 41.33 & $+$33 25 18.26 & $83^{5}$ & 6.27 & 8.30 & 8.72 & 6.55 & 42.0 & 13.8 & 3.71\\
NGC 3003 & 09 48 36.05 & $+$33 25 17.40 & $85^{2}$ & 25.4 & 22.16 & 10.03 & 30.16 & 79.0 & 13.1 & 4.14\\
NGC 3044 & 09 53 40.88 & $+$01 34 46.70 & $85^{2,6}$ & 20.3 & 12.99 & 9.56 & 17.60 & $-$66.5 & 15.8 & 2.76\\
NGC 3079 & 10 01 57.80 & $+$55 40 47.24 & $84^{2}$ & 20.6 & 23.07 & 10.00 & 29.40 & $-$12.5 & 13.9 & 5.49\\
NGC 3556 & 11 11 30.97 & $+$55 40 26.80 & $81^{1}$ & 14.09 & 15.98 & 9.68 & 20.22 & 81.5 & 12.9 & 5.12\\
NGC 3877 & 11 46 07.70 & $+$47 29 39.65 & $85^{1}$ & 17.7 & 13.13 & 9.17 & 11.07 & 34.5 & 13.6 & 4.33\\
NGC 4096 & 12 06 01.13 & $+$47 28 42.40 & $82^{1}$ & 10.32 & 9.61 & 9.16 & 10.94 & 20.0 & 14.6 & 3.91\\
NGC 4157 & 12 11 04.37 & $+$50 29 04.80 & $83^{2,7}$ & 15.6 & 15.88 & 9.72 & 21.22 & 65.0 & 13.9 & 3.97\\
NGC 4217 & 12 15 50.90 & $+$47 05 30.40 & $86^{2,8}$ & 20.6 & 15.28 & 9.44 & 15.29 & 49.3 & 14.7 & 3.33\\
NGC 4302 & 12 21 42.48 & $+$14 35 53.90 & $90^{2,6}$ & 19.41 & 13.27 & 9.24 & 12.06 & $-$0.2 & 15.8 & 2.98\\
NGC 4565 & 12 36 20.78 & $+$25 59 15.63 & $86^{3,7}$ & 11.9 & 28.04 & 9.80 & 23.14 & $-$44.3 & 14.6 & 3.23\\
NGC 4631 & 12 42 08.01 & $+$32 32 29.40 & $89^{4}$ & 7.4 & 15.82 & 9.33 & 13.45 & 85.0 & 14.2 & 4.54\\
NGC 5084 & 13 20 16.92 & $-$21 49 39.30 & $90^{1}$ & 23.4 & 42.54 & 9.96 & 27.98 & 75.5 & 25.3 & 1.99\\
NGC 5775 & 14 53 57.60 & $+$03 32 40.05 & $86^{2}$ & 28.9 & 16.39 & 10.09 & 32.82 & $-$33.7 & 21.3 & 1.93\\
UGC 10288 & 16 14 24.80 & $-$00 12 27.10 & $90^{2}$  & 34.1 & 24.30 & 9.98 & 28.63 & $-$89.9 & 18.5 & 2.46\\
  \noalign{\smallskip}\hline
\end{tabular}
\ec
\tablecomments{0.99\textwidth}{$^a$ From the NASA Extragalactic Database (NED).\\
$^b$ Taken from Ref--- 1. Paper I \citep{2012AJ....144...43I} 2. Paper IX \citep{2018AA...611A..72K}. 3. Paper XVI \citep{schmidt2019chang}. 4. Paper XIV \citep{mora2019chang}. 5. \cite{Vollmer2016TheFH}. 6. \cite{2015ApJ...799...61Z}. 7. \cite{2014AJ....148..127Y}. 8. \cite{Allaert2015HERschelOO}. \\
$^c$ Taken from Paper IV \citep{wiegert2015chang}. \\
$^d$ Observed blue radius at the 25th mag arcsec$^{-2}$ isophote taken from Paper I \citep{2012AJ....144...43I}.\\
$^e$ The $\hi$ mass and radius from CHANG-ES paper XXVI (Zheng et al. in Prep). The $\hi$ radius is calculated by the $\hi$ size-mass relation from \cite{2016MNRAS.460.2143W}. \\
$^f$ The position angle of $\hi$ disk (see section 3.1).\\
$^g$ The information of $\hi$ intensity maps.\\}
\end{table}

\section{Deriving the HI scale height}

We use a photometric approach instead of kinematic approach to determine scale heights due to two characteristics of our sample. Firstly the accuracy of 3-D kinematic modeling routines is challenged by highly inclined observations. Secondly, the low-velocity resolution of the CHANG-ES $\hi$ cubes doesn't permit accurate modeling.

We firstly derive the raw photometric $\hi$ scale height ($h_{\rm phot}$) in a similar way as in Z21 (see section 3.1), but skip their relatively simple correction for beam smearing and planar projection (caused by the not perfectly edge-on inclinations of galaxies). Instead, we investigate in detail and design a more robust procedure to correct for systematic biases including these two effects in Z21 and additional edge-on projection effect (mainly caused by flaring).

\subsection{Deriving the raw photometric scale heights}

The raw photometric scale heights $h_{\rm phot}$ are derived in two steps: deriving the vertical profile of surface densities in strips perpendicular to the disk mid-plane, and fitting the vertical distribution profiles with a Gaussian function to obtain the width \footnote{In Z21, to direct compare with the radio continuum scale height, we fit an exponential function. Changing from an exponential function to a Gaussian function here improves the fitting result as the median of reduced $\chi^2$ decreases from 7.61 to 6.16, and the major trends presented in Z21 do not significantly change, which we demonstrate later in Figure \ref{figure:resultmain} and \ref{figure:resultsfr}.}. 

These steps are similar to the “BoxModels” task in the new NOD3 program package \citep{2017A&A...606A..41M}. \citet{2018AA...611A..72K} applied the “BoxModels” task to measure the radio halo scale height of CHANG-ES sample. However, the $\hi$ disks are thin and asymmetric compared with the radio halos. Z21 thus designed additional steps specifically for measuring $\hi$ scale heights. They derived the position angle of the $\hi$ plane instead of using the optical position angle. They also measured the vertical distributions on both sides separately. 

As in Z21, we only use the radially averaged $h_{\rm phot}$ within the optical radius $R_{25}$ ($\bar{h}_{\rm phot}$) in the analysis.
This is because the nearly edge-on view prevents us from recovering the flaring shape of the radial profile of scale-heights.
$\bar{h}_{\rm phot}$ should thus be only viewed as an indicator of the $\hi$ disk thickness. 
The error of each $\bar{h}_{\rm phot}$ combines the uncertainty from the model fitting and the scatter of $h_{\rm phot}$ in the profile. 

\begin{figure}
  \centering
  \includegraphics[width=150mm]{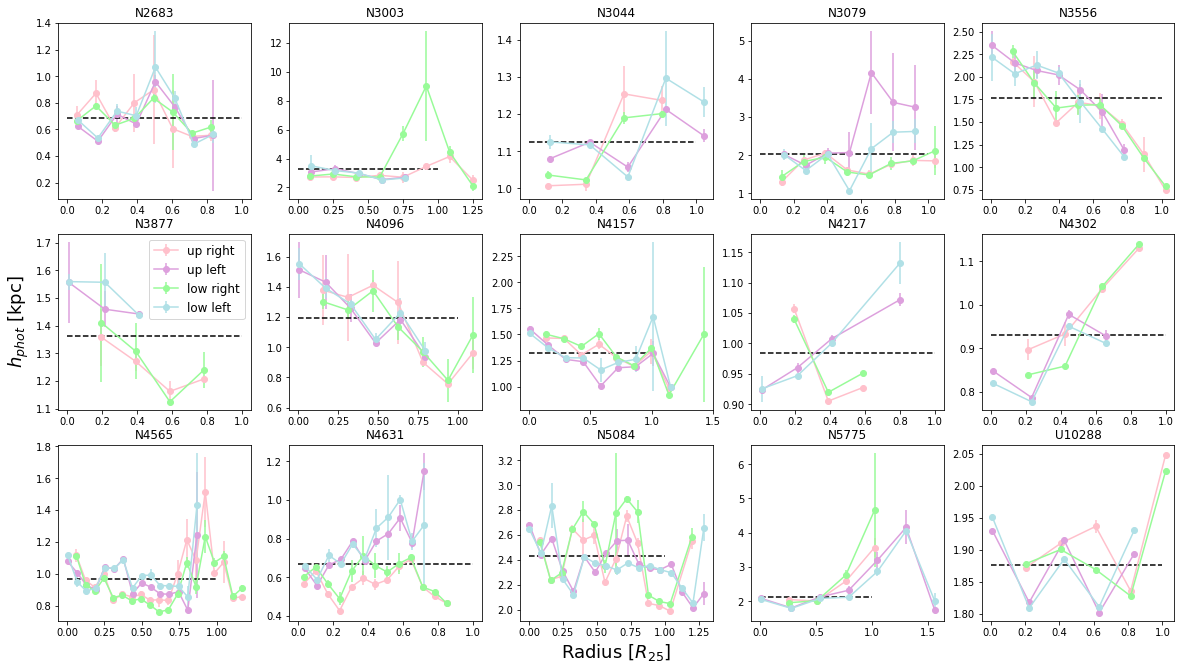}
\caption{The radial profile of $h_{\rm phot}$. The black dash lines represent the $\bar{h}_{\rm phot}$. The profiles in warm colors (pink and purple) and profiles in cool colors (green and blue) represent the results of two sides respect to the galactic mid-plane (`x-axis'), which are labeled by `up' and `low', respectively. The profiles in dark colors (purple and blue) and profiles in light colors (pink and green) represent the results of two sides respect to minor axis (`y-axis'), which are labeled by `left' and `right', respectively.}
\label{fig:profile}
\end{figure}

Figure \ref{fig:profile} shows the radial profile of $h_{\rm phot}$ and the value of $\bar{h}_{\rm phot}$ (black dash line) in each galaxy. We present four profiles in each galaxy. The fitting results of the `up' and `low' represent two sides with respect to the mid-plane of a disk, while the `left' and `right' are two sides with respect to the minor axis of a disk. The `up' and `low' profiles here correspond to (but not the same, as the deriving methods are different) the red and blue fitting curves in Figure A2 of Z21. We show these profiles mainly to demonstrate how $\bar{h}_{\rm phot}$ are derived, but emphasize that only $\bar{h}_{\rm phot}$ are considered reliable in the following analysis.

\subsection{New procedure to correct for artificial broadening}

\cite{2018AA...611A..72K} and Z21 mainly considered the beam smearing effect and planar projection effect, which have artificially increased the measured scale-heights. For galaxies with an inclination lower than $90^{\circ}$, the flux distribution away from the mid-plane is considered to be a mixture of the intrinsic vertical component and the projected planar component of fluxes from the disk. In the observed data, such a mixture is further convolved with the beam PSF. These smearing effects need be removed from $h_{\rm phot}$ before we obtain the final measurements of $\hi$ scale height. As we describe below, we correct for these effects in different ways from that of \cite{2018AA...611A..72K} and Z21.

In Z21, the contamination of the planar projection was accounted for as a pseudo increment of the PSF FWHM along the z direction in a radially dependent way: $\Delta {\rm FWHM} = R \ {\rm cos}(r/R* \pi/2)  {\rm cos} \ i$, where $R$ was the disk size, $r$ was the radius, and $i$ was the inclination. The broadened PSF then had an effective FWHM along the z direction : ${\rm FWHM}_{\rm eff,z} = \sqrt{b_{maj}^2+\Delta {\rm FWHM}^2}$. In Z21, this effective PSF was convolved with the strip profile model before being compared with the data, so the best-fit scale-heights were expected to be clean from those contaminating effects. Our new procedure to correct for these two effects does not assume the effective FWHM. We firstly derive the photometric scale-heights without corrections, and then apply correction equations which are calibrated by comparing the real and measured values of scale-heights from moment-0 images of mock cubes. The details are described in section 3.3. 

Additionally, we consider a third systematic effect artificially increasing $\bar{h}_{\rm phot}$ but not considered in Z21, called edge-on projection. This effect is the projection of emission from outer disks even when the disk is perfectly edge-on, as most disks have the flaring feature \citep{brinks1984high, bigiel2012universal}. Such an effect is exacerbated in the $\hi$ observations as the $\hi$ disks are flat and extended in $\hi$-rich galaxies \citep{2008AJ....136.2563W, 2008AJ....136.2648D}. The correction for this edge-on projection effect on the photometric scale-heights is presented in section 3.3. 

\subsection{Investigation of scale-height broadening based on mock cubes}

We use mock cubes to quantify the extent of over-estimating the scale-heights due to effects of PSF smearing, planar projection, and edge-on projection. The errors are calculated as 1-$\sigma$.

We generate mock $\hi$ cubes with the GALMOD task of BBarolo \citep{2015MNRAS.451.3021D}, using the $\hi$ scale height profiles derived in \cite{2019A&A...622A..64B} for eight THINGS galaxies, along with the velocity dispersion profiles, surface density profiles, and rotation curves in that paper. We have excluded two galaxies from the original sample of \cite{2019A&A...622A..64B}, DDO 154 and NGC 2976, because of their low stellar masses ${\rm log} M_{*} =$ 7.1 and 9.1, while the lowest stellar mass of the CHANG-ES edge-on sample is ${\rm log} M_{*} =$ 9.96. We call these eight galaxies the input sample.

The inclination $i$ and maximum radius of ring $r_{\rm max}$ (larger than $R_{25}$ in all cases) are fixed for each model disk, but vary between model disks. Each model disk is built with rings of different radius $r$. For each $r$, the $\hi$ ring has scale height $h_r$, rotation velocity $v_r$, velocity dispersion $\sigma_r$, and surface density $\Sigma_r$, with values determined by interpolating the related radial profiles of a galaxy from the input sample. The last data point in a radial profile is repeated if $r_{\rm max}$ exceeds that. The cube and resulted moment-0 map are firstly generated at the original THINGS resolution (i.e. the highest resolution) with GALMOD. Later, the spatial resolution of the output moment-0 maps can be changed through convolving with gaussian kernels. We add gaussian noise with sigma equivalent to the median rms of the CHANG-ES HI moment-0 maps to the model moment-0 maps. Similar to how we treat the CHANG-ES data, when deriving $h_{\rm phot}$, we only use the data points with signal to noise ratio larger than 3 along each vertical strip to fit gaussian models. We use the scale-height averaged within the radius R$_{25}$, $\bar{h}$, as the parameter to be tested. We compare the photometrically measured $\bar{h}$ ($\bar{h}_{\rm phot}$) to the real $\bar{h}$ of models ($\bar{h}_{\rm true}$). 

In the following, we firstly investigate separately each of the effects causing systematic biases, including PSF smearing, planar projection, and edge-on projection. Then we take these effects together, and correct $\bar{h}_{\rm phot}$ against $\bar{h}_{\rm true}$.

\subsubsection{Effect of PSF smearing}

\begin{figure}
  \centering
   \includegraphics[width=0.65\linewidth]{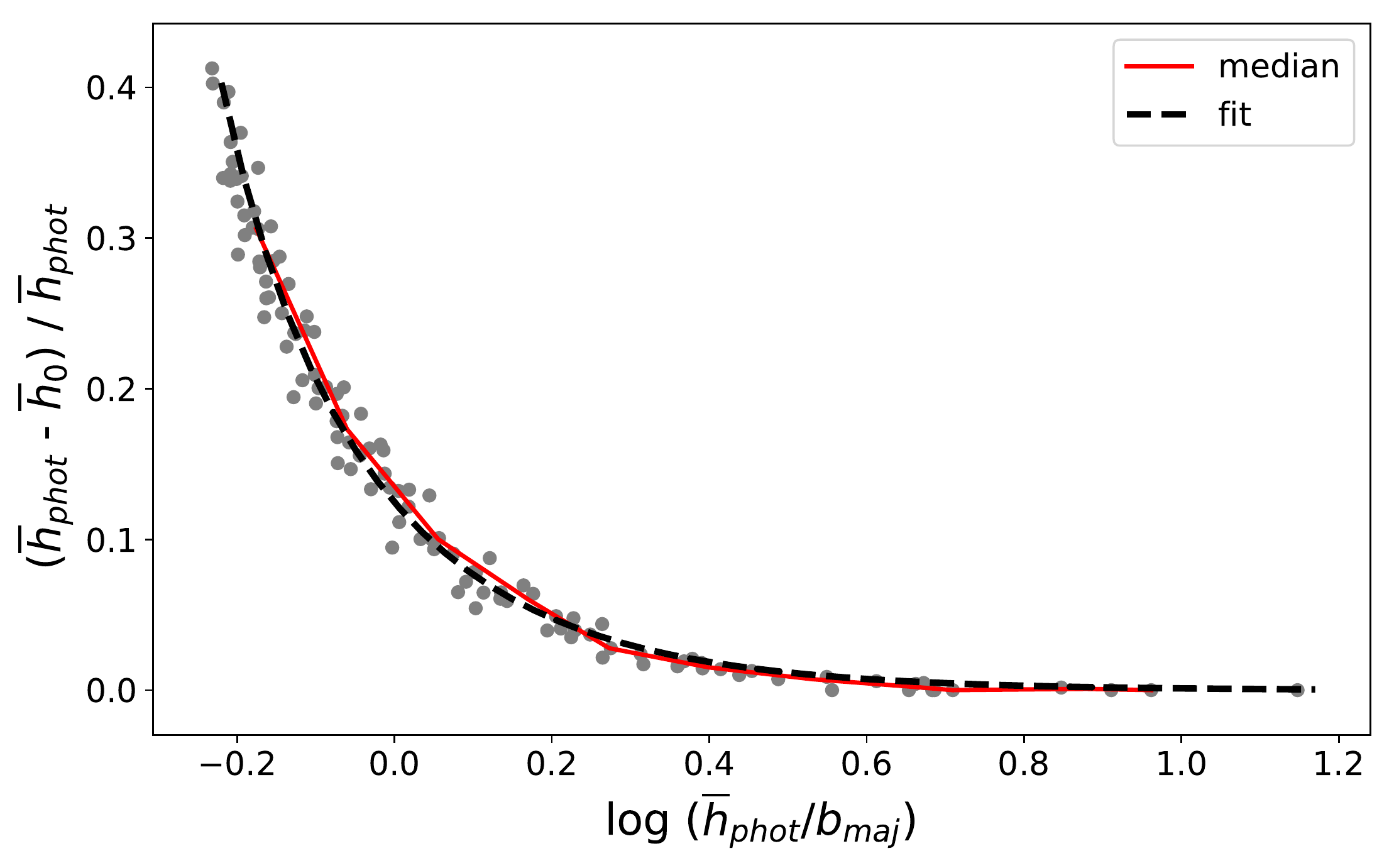}
\caption{The PSF smearing on photometric scale-height $\bar{h}_{\rm phot}$, with the best-fit (black dash) line and median distribution (red line).}
\label{PSFsmearing}
\end{figure}

For the test of the PSF smearing effect, we fix the inclination of the mock galaxies to 90$^{\circ}$ and vary the beam's major axis ($b_{maj}$) of the data cube from the highest resolution of $b_{maj}=1.5''$ (pixel size of the THINGS data) to the lowest resolution corresponding to the minimum value of the uncorrected $\bar{h}_{\rm phot}/b_{maj}$ in the CHANG-ES edge-on sample, with a step of $\Delta b_{maj}=1.5''$. We generate 117 mock $\hi$ cubes. We quantify with decreasing resolution, how $\bar{h}_{\rm phot}$ is increasingly over-estimated with respect to the measurement at the best resolution $\bar{h}_{0}$. The result is shown in Figure \ref{PSFsmearing}. The median trend of the over-estimation has a relatively small scatter of 1.85\%, and is fitted with the following equation: 
\begin{equation}
\bar{h}_{\rm phot}^{2}-\bar{h}_{0}^2 = 0.234 \ b_{maj}^2,
\end{equation}

which could be approximated as a convolution with an effective Gaussian kernel. When the beam is less than 25\% of $\bar{h}_{\rm phot}$, the $\bar{h}_{\rm phot}$ will have an uncertainty due to PSF smearing of 0.19\%.

\begin{figure}
\includegraphics[width=0.49\linewidth]{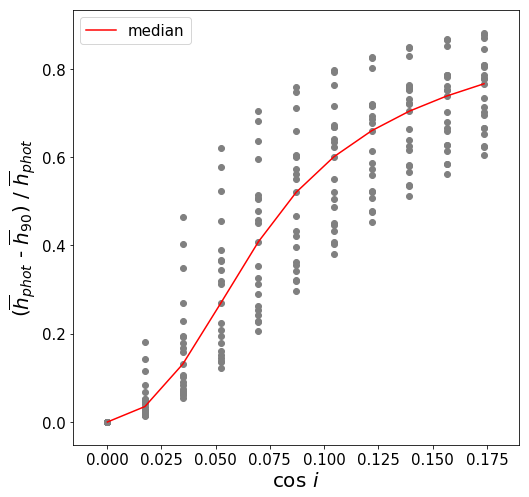}
\includegraphics[width=0.49\linewidth]{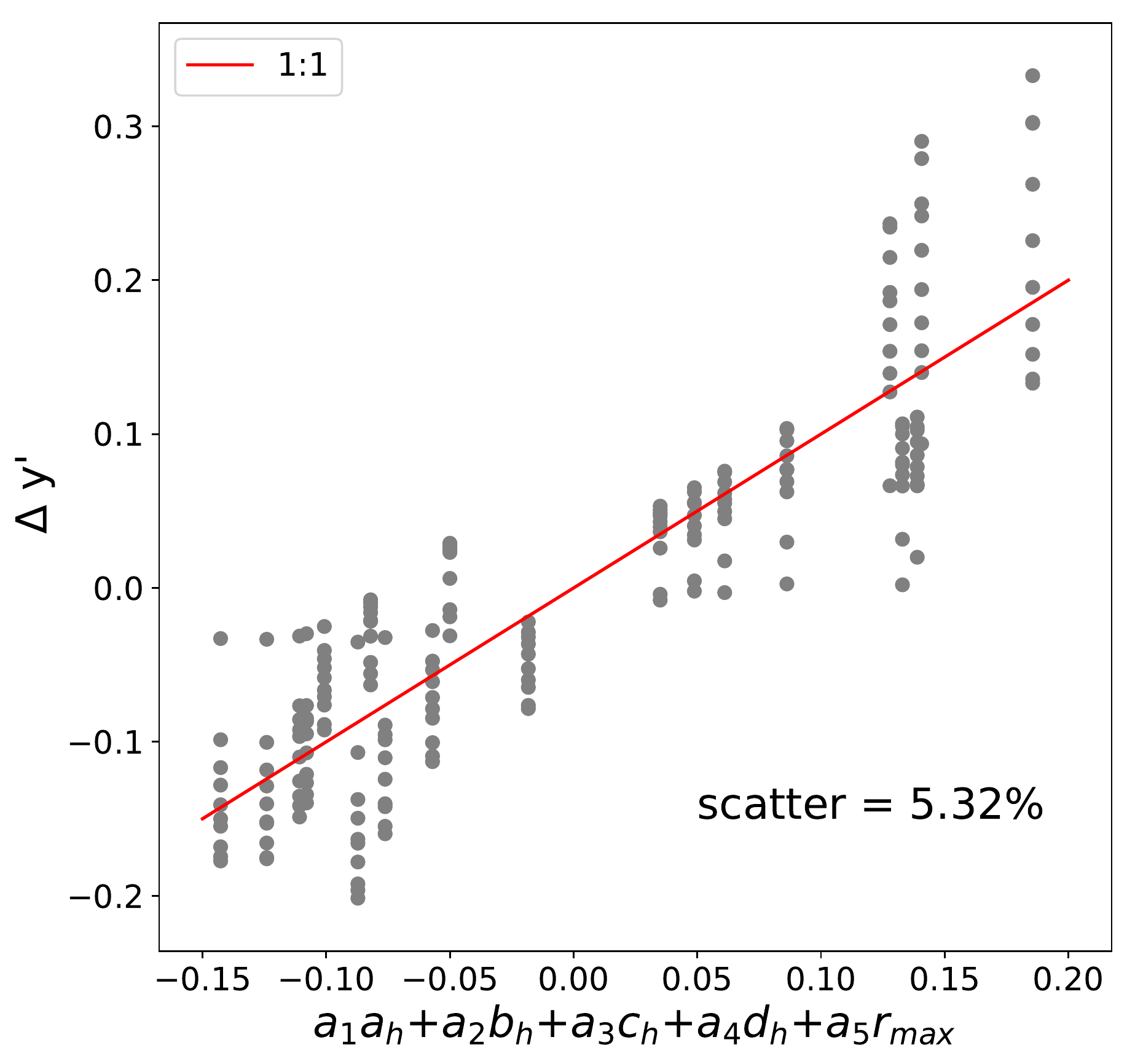}
\begin{center}
{\textbf{(a) \hspace{65mm} (b)}}
\end{center}
\includegraphics[width=0.49\linewidth]{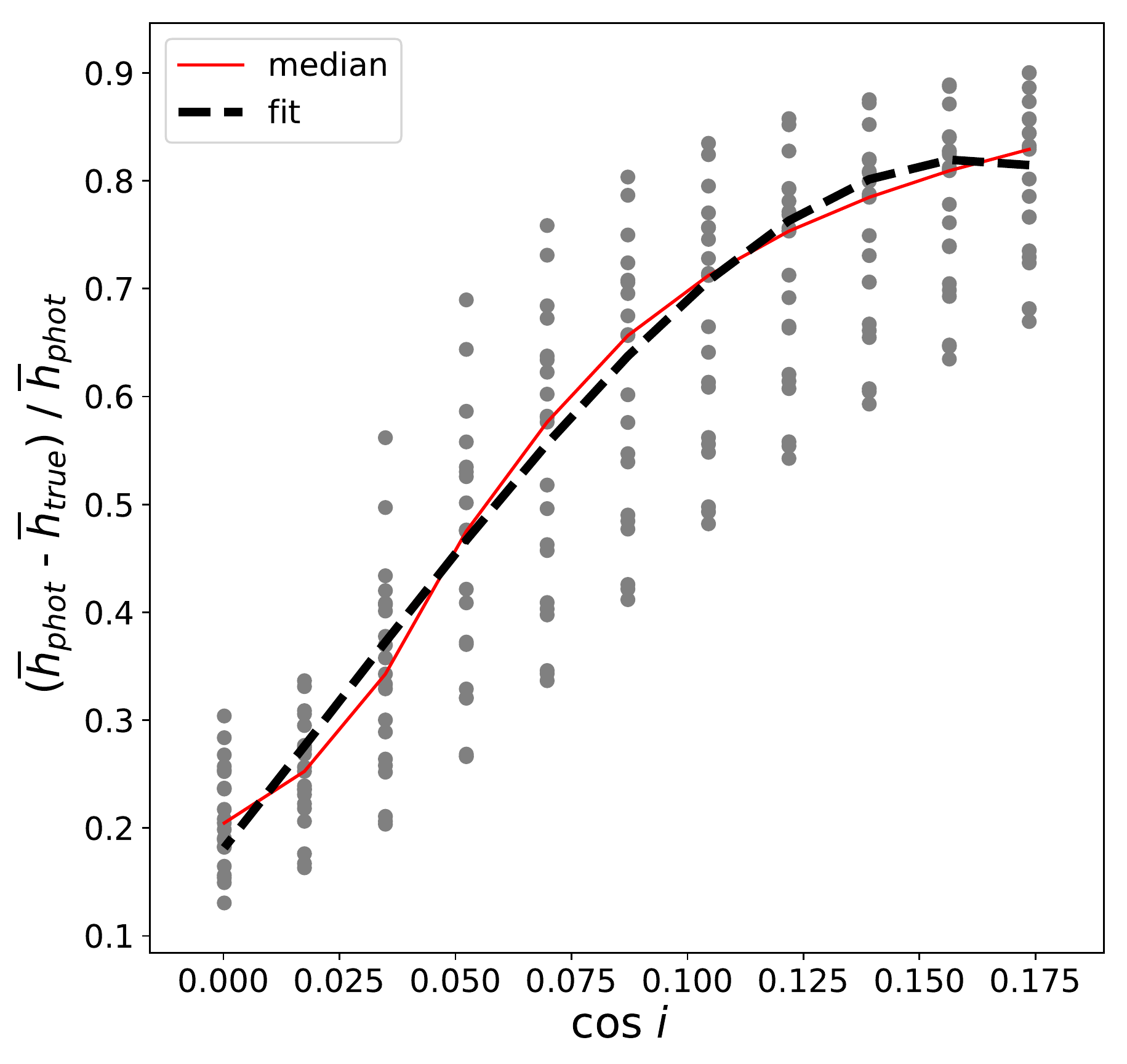}
\includegraphics[width=0.49\linewidth]{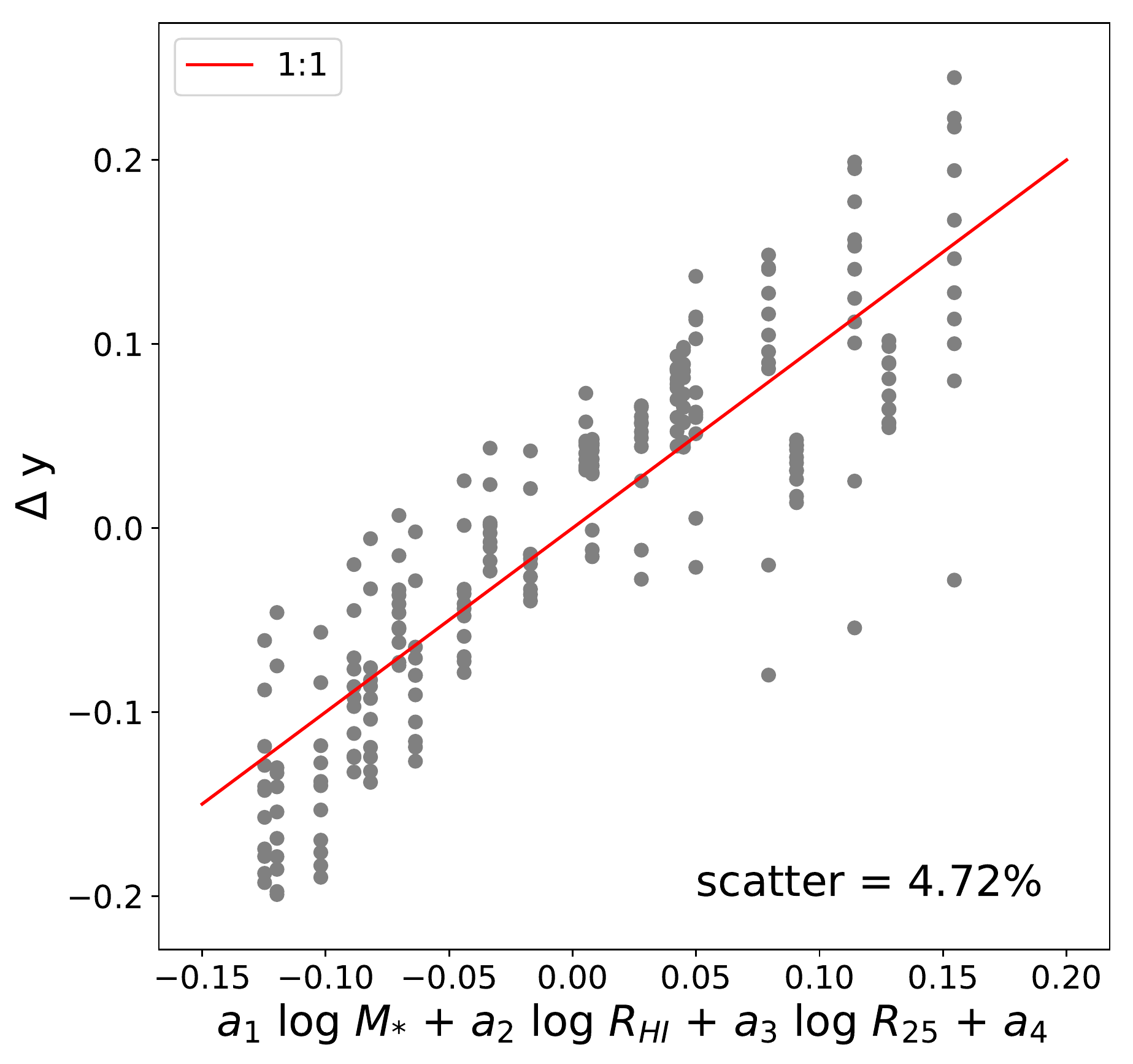}
\begin{center}
{\textbf{(c) \hspace{65mm} (d)}}
\end{center}
\caption{Mock tests about planar and edge-on projections. (a) The planar projection effect: the relation between ${\rm y'} = (\bar{h}_{\rm phot}-\bar{h}_{90})/\bar{h}_{\rm phot}$ and $\cos i$. The red curve is the median relation. (b) The effect of edge-on projection in causing the scatter in panel a: the relation between $\Delta {\rm y'}$ and parameters describing the shape of flaring of disk. The red line is 1:1 line. (c) Similar to panel a, but $\bar{h}_{90}$ is replaced by $\bar{h}_{true}$. The black dashed curve is best-fit 3rd order polynomial relation to the red curve. (d) Similar as panel a, but $\Delta {\rm y}$ is the offset of data points from the best-fit relation (black dashed curve) in panel c.}
\label{planarprojection1}
\end{figure}

\subsubsection{Effect of planar projection}

In the following, we investigate the planar projection effect, which however is likely to interfere with the edge-on projection effect. We use the highest resolution of $1.5''$, and vary the inclination of the galaxy disk from 80$^{\circ}$ to 90$^{\circ}$ with a step of 1 degree. We also vary the disk sizes in the mocks, $r_{max}$. The range of $r_{max}$ is set such that $r_{max}/R_{25}$ are between 1 and maximum $R_{\rm HI}/R_{25}$ of CHANG-ES $\hi$ sample. We build 21 mock cubes for each inclination, thus in total 231 mock cubes.

We quantify with decreasing inclination angle, how $\bar{h}_{\rm phot}$ is increasingly over-estimated with respect to the measurement when the galaxy is perfectly edge-on $\bar{h}_{90}$. Figure \ref{planarprojection1}(a) shows the result of the 231 mocks (points), with the median distribution (red line). At a given inclination, the planar projection effect has large scatter varying between different input galaxies. We explain and justify below, the edge-on projection is likely responsible for this large scatter.

\subsubsection{Effect of edge-on projection}

The edge-on projection is mainly caused by disk flaring. It interferes with the planar projection because each line of sight intercept different (and more) disk rings in a flared disk from in a flat disk.

To parametrize the shape of underlying $\hi$ flaring, for each galaxy in the input sample, we fit the scale height radial profile with a 3-rd order polynomial equation $h(r) = a_hr^3+b_hr^2+c_hr+d_h$. The 4 coefficients of the polynomial equation, together with $r_{max}$, are parameters to quantify the effect of edge-on projection. We 
define ${\rm y'} = (\bar{h}_{\rm phot}-\bar{h}_{90})/\bar{h}_{\rm phot}$, so that the scatter of data points from the median curve (the red line in Figure \ref{planarprojection1}(a)) is quantified as $\rm \Delta y' = y'-median(y')$. We fit a linear relation of the 5 parameters describing the edge-on projection to predict $\rm \Delta y'$ (dashed line in Figure \ref{planarprojection1}(b)). The strong linear correlation with a small scatter of 5.32\% implies that the effect of edge-on projection indeed accounts for a large fraction of the scatter around the median relation in Figure \ref{planarprojection1}(a).

The analysis in Figure \ref{planarprojection1}(a) and (b) proves that the edge-on projection effect is indeed involved in the planar projection effect. The underlying flaring shape and $r_{max}$ used to fit $\rm \Delta y'$ in \ref{planarprojection1}(b) are inputs of mocks which cannot be directly measured in observations. Motivated by the quasi-static equilibrium model of gas (based on which \cite{2019A&A...622A..64B} derived the scale-heights of the input sample), we consider $M_*$, $M_{\rm HI}$, $R_{\rm HI}$, $R_{25}$ and SFR as candidate parameters that may mimic the combined effect of the underlying shape of $\hi$ flaring and $r_{max}$. We test different combination of these candidate parameters and find that the combination of $M_*$, $R_{\rm HI}$ and $R_{25}$ are enough to derive similarly tight relation with $\rm \Delta y'$ as when using $r_{max}$ and the 4 coefficients describing the underlying shape of $\hi$ flaring.

In addition to interfering with planar projection, the edge-on projection further cause a difference between the perfectly edge-on $\bar{h}_{90}$ and model input $\bar{h}_{true}$. This because in an edge-on view, the flaring outer disk can dominate the surface brightness at high $z$, where $z$ is the vertical distance from the mid-plane. 
Thus, in the following, we consider the planar projection and edge-on projection effects together, and correct $\bar{h}_{\rm phot}$ to $\bar{h}_{true}$.

\subsubsection{Correcting for both planar and edge-on projection effects together}
We correct two projection effects from $\bar{h}_{\rm phot}$ to derive $\bar{h}_{true}$. We do it in a similar way as in the last section, with two major modifications: $\bar{h}_{90}$ is replaced by $\bar{h}_{true}$, and the parameters describing the flaring profile shapes are replaced by $M_*$, $R_{\rm HI}$ and $M_{\rm HI}$.

The median relation between $(\bar{h}_{\rm phot}-\bar{h}_{true})/\bar{h}_{\rm phot}$ and $\cos i$ is derived to quantify the effect of planar projection (Figure \ref{planarprojection1}(c)). 
The median trend has a scatter of 10.04\%. Interestingly, the scatter looks much smaller than in Figure \ref{planarprojection1}(a), possibly because different systematic biases cancel out. A 3-rd order polynomial equation is fitted to the median curve:
\begin{equation}
    \frac{\bar{h}_{\rm phot}-\bar{h}_{true}}{\bar{h}_{\rm phot}} = c_3(cos \ i)^3 + c_2(cos \ i)^2 + c_1(cos \ i) + c_0,
\end{equation}
in which $c_3 = -97.91$, $c_2 = 7.15$, $c_1 = 5.36$ and $c_0 = 0.18$.

We calculate the offset of $(\bar{h}_{\rm phot}-\bar{h}_{true})/\bar{h}_{\rm phot}$ from the prediction of the best-fit relation, and denote it as $\rm \Delta y$. Finally, we fit a linear relation of $\Delta \rm y$ as a function of $M_*$, $R_{\rm HI}$ and $R_{25}$. 

\begin{equation}
    \Delta {\rm y} = 0.28 ~ \log M_*+ 0.68 ~ \log R_{\rm HI}-0.99 ~ \log R_{25}-2.69
\end{equation}

\begin{figure}
   \centering
   \includegraphics[width=0.5\linewidth]{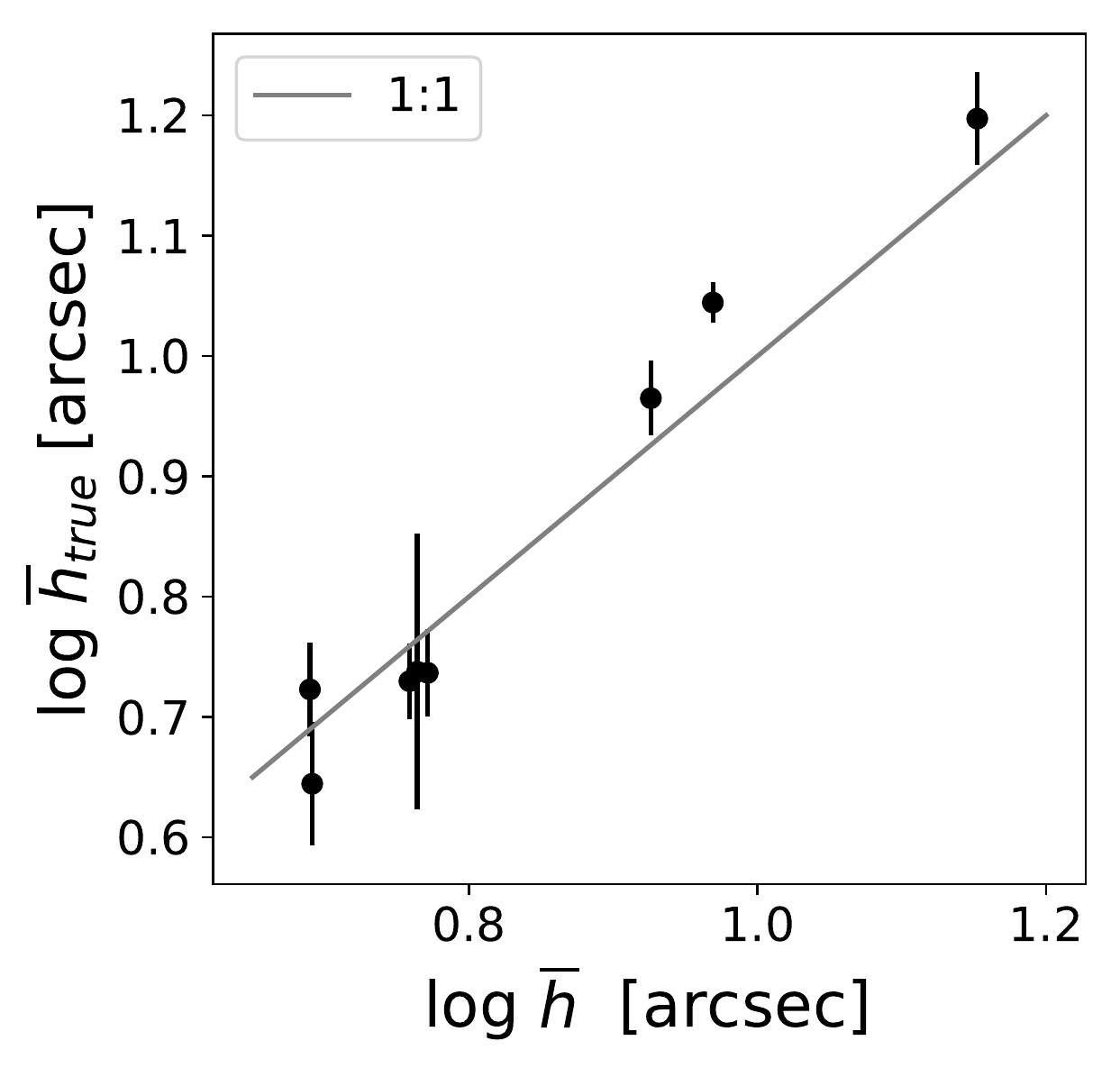}
\caption{The corrected scale-heights $\bar{h}$ of mock cubes with CHANG-ES edge-on sample median parameters vs. input value $\bar{h}_{\rm true}$. Grey line is 1:1 line.}
\label{figure:edgeon}
\end{figure}

We use the equations (1-3) derived above, to correct for the three associated observational effects together from photometric scale-heights.
To test the goodness of such a methodology, we produce mock cubes with the median beam size (15 arcsec) and median inclination angle (85 degree) of the CHANG-ES edge-on sample. We obtain the photometric scale-heights, and correct for the observational effects as described above. Figure \ref{figure:edgeon} shows that the corrected scale-heights agree very well with the true scale-heights. The fit shows that the uncertainty in the photometrically measured scale height is typically 9.36\% of the true scale height, when the mock galaxies have median properties of CHANG-ES galaxies.

We do not consider the projection effect of warps, for with the limited sensitivity and spectral resolution, as well as the nearly edge-on nature of the sample, it is difficulty to reliably identify warps misaligned with respect to the inclination of the main disk. For example, most warps start near or beyond $R_{25}$ \citep{1990ApJ...352...15B} but due to our limited sensitivity we are unable to detect much $\hi$ beyond $R_{25}$ (median $R_{\rm HI}/R_{25} = 1.14 \pm 0.33$). Fortunately, this may reduce a putative warp’s contribution to our photometric determination of $\bar{h}_{\rm phot}$. Additionally the potential contamination from warps is difficult to quantify even when kinematical modeling is involved \citep{2014AJ....148..127Y, 2019A&A...622A..64B}. For these reasons we leave the effect of warps as a caveat to be investigated in the future.

In summary, these mock tests justify that photometrically derived scale-heights can indicate the true scale-heights of galaxies after properly account for the observational effects of PSF smearing, planar projection, and edge-on projection. 

\subsection{Deriving the corrected scale-heights}
We apply the correction equations estimated from the mock tests to the CHANG-ES photometric measurements $\bar{h}_{\rm phot}$, and calculate the corrected scale-heights $\bar{h}$, which are presented in Table \ref{table3:hHI}. The final errors of $\bar{h}$ in this table include the uncertainties associated with these corrections and the propagated uncertainty of the disk inclination ($\pm1^{\circ}$). The scientific analysis will be conducted based on these corrected values. We note that the galaxies largely effected by beam smearing are NGC 4302, NGC 5084, and UGC 10288 due to their low resolution compared with uncorrected photometric scale height, while the uncertainties in Table \ref{table3:hHI} for NGC 3877 and NGC 4157 are dominated by inclination angle. The edge-on projection correction significantly reduces the putative overestimate of the photometric scale height. After these corrections NGC 5084 whose $\hi$ distribute in a large ring, and the tidally perturbed galaxy NGC 3003 which has the lowest mass density among the sample, still have large $\bar{h}$, while the remainder of the galaxies have scale heights approaching those in the literature \citep{2019A&A...622A..64B}.

\begin{table}
\bc
\begin{minipage}[]{100mm}
\caption[]{The $\hi$ Scale Height and Other Physical Parameters\label{table3:hHI}}\end{minipage}
\setlength{\tabcolsep}{1pt}
\small
 \begin{tabular}{ccccccccccc}
 \hline\noalign{\smallskip}
Galaxies & $\bar{h}_{\rm phot}$ & $\bar{h}$ & $\bar{h}$ & $\bar{h}^a_{\rm Z21}$ & $\Sigma_{\rm Mtot,r25}^b$ & $\Sigma_{\rm Mbaryon,r25}^c$ & SFR$^d_{\rm H_{\alpha}+22\mu m}$\\
 & [$''$] & [$''$] & [kpc] & [kpc] & [$10^7 M_{\odot}\rm kpc^{-2}$] & [$10^7 M_{\odot}\rm kpc^{-2}$] & [$M_{\odot}\rm yr^{-1}$]\\
  \hline\noalign{\smallskip}
NGC 2683 & ~ 22.64$\pm$4.60 ~ & ~ 9.69$\pm$2.18 ~ & ~ 0.29$\pm$0.07 ~ & ~ 0.79$\pm$0.27 & 29.84 & 6.70 & 0.25 $\pm$ 0.03\\
NGC 3003 & ~ 26.77$\pm$11.53 ~ & ~ 13.00$\pm$5.94 ~ & ~ 1.60$\pm$0.73 ~ & ~ 4.33$\pm$2.08 & 3.64 & 1.69 & 1.56 $\pm$ 0.16\\
NGC 3044 & ~ 11.43$\pm$0.94 ~ & ~ 4.00$\pm$0.63 ~ & ~ 0.39$\pm$0.06 ~ & ~ 1.0$\pm$0.28 & 12.18 & 3.11 & 1.75 $\pm$ 0.16\\
NGC 3079 & ~ 20.26$\pm$6.59 ~ & ~ 6.82$\pm$2.52 ~ & ~ 0.68$\pm$0.25 ~ & ~ 2.35$\pm$1.15 & 7.01 & 4.58 & 5.08 $\pm$ 0.45\\
NGC 3556 & ~ 25.82$\pm$5.29 ~ & ~ 4.91$\pm$1.12 ~ & ~ 0.34$\pm$0.08 ~ & ~ 1.89$\pm$0.45 & 7.51 & 5.15 & 3.57 $\pm$ 0.3\\
NGC 3877 & ~ 15.90$\pm$1.69 ~ & ~ 7.56$\pm$1.00 ~ & ~ 0.65$\pm$0.09 ~ & ~ 1.74$\pm$0.2 & 10.15 & 5.77 & 1.35 $\pm$ 0.12\\
NGC 4096 & ~ 23.93$\pm$4.43 ~ & ~ 8.28$\pm$1.71 ~ & ~ 0.41$\pm$0.09 ~ & ~ 1.48$\pm$0.28 & 13.42 & 3.82 & 0.71 $\pm$ 0.08\\
NGC 4157 & ~ 17.51$\pm$1.70 ~ & ~ 5.02$\pm$0.64 ~ & ~ 0.38$\pm$0.05 ~ & ~ 1.23$\pm$0.28 & 12.62 & 4.90 & 1.76 $\pm$ 0.18 \\
NGC 4217 & ~ 9.85$\pm$0.67 ~ & ~ 3.66$\pm$0.56 ~ & ~ 0.37$\pm$0.06 ~ & ~ 0.93$\pm$0.17 & 14.61 & 5.15 & 1.89 $\pm$ 0.18\\
NGC 4302 & ~ 9.89$\pm$1.16 ~ & ~ 6.03$\pm$1.76 ~ & ~ 0.57$\pm$0.17 ~ & ~ 0.96$\pm$0.21 & 14.36 & 4.16 & 0.92 $\pm$ 0.08\\
NGC 4565 & ~ 16.76$\pm$2.51 ~ & ~ 9.56$\pm$1.76 ~ & ~ 0.55$\pm$0.10 ~ & ~ 1.27$\pm$0.22 & 11.15 & 3.03 & 0.96 $\pm$ 0.09\\
NGC 4631 & ~ 18.66$\pm$4.01 ~ & ~ 17.85$\pm$4.45 ~ & ~ 0.64$\pm$0.16 ~ & ~ 0.78$\pm$0.27 & 6.99 & 2.47 & 2.62 $\pm$ 0.22\\
NGC 5084 & ~ 21.45$\pm$1.89 ~ & ~ 18.47$\pm$2.43 ~ & ~ 2.10$\pm$0.28 ~ & ~ 2.95$\pm$0.35 & 14.30 & 1.98 & 0.1 $\pm$ 0.03$^*$\\
NGC 5775 & ~ 15.17$\pm$1.86 ~ & ~ 3.14$\pm$0.75 ~ & ~ 0.44$\pm$0.11 ~ & ~ 2.01$\pm$0.56 & 14.57 & 9.31 & 7.56 $\pm$ 0.65 \\
UGC 10288  & ~ 11.35$\pm$0.29 ~ & ~ 7.19$\pm$0.51 ~ & ~ 1.19$\pm$0.09 ~ & ~ 1.91$\pm$0.09 & 6.33 & 1.87 & 0.66 $\pm$ 0.07\\  \noalign{\smallskip}\hline
\end{tabular}
\ec
\tablecomments{0.99\textwidth}{
$^a$ The Gaussian averaged scale height through corrected using the method in Z21.\\
$^b$ Total mass surface density $\Sigma_{\rm Mtot,r25} = M_{\rm tot} / (\pi* R_{25}^2)$, where $M_{tot}$ \citep{2012AJ....144...43I} is the total mass within the blue isophotol radius $R_{25}$ in kpc, calculated using the velocity amplitude based on line width. \\
$^c$ Baryon mass surface density $\Sigma_{\rm Mbaryon,r25} = M_{\rm baryon} / (\pi* R_{25}^2)$, where $M_{\rm baryon} = 1.4 \times M_{\rm HI}+M_{*}$  is the baryon mass within the blue isophotol radius $R_{25}$ in kpc.\\
$^d$ Star formation rates were estimated by combination of H$\alpha$ and 22 $\mu$m data, taken from Paper XVII \citep{2019ApJ...881...26V}, except NGC 5084. Lacking H$\alpha$ observation, the star-formation rate of NGC 5084 is only estimated by 22 $\mu$m data, taken from Paper IV \citep{wiegert2015chang}.\\
}
\end{table}

\section{Results}

\begin{figure}
  \begin{minipage}[t]{0.99\linewidth}
  \centering
   \includegraphics[width=0.45\linewidth]{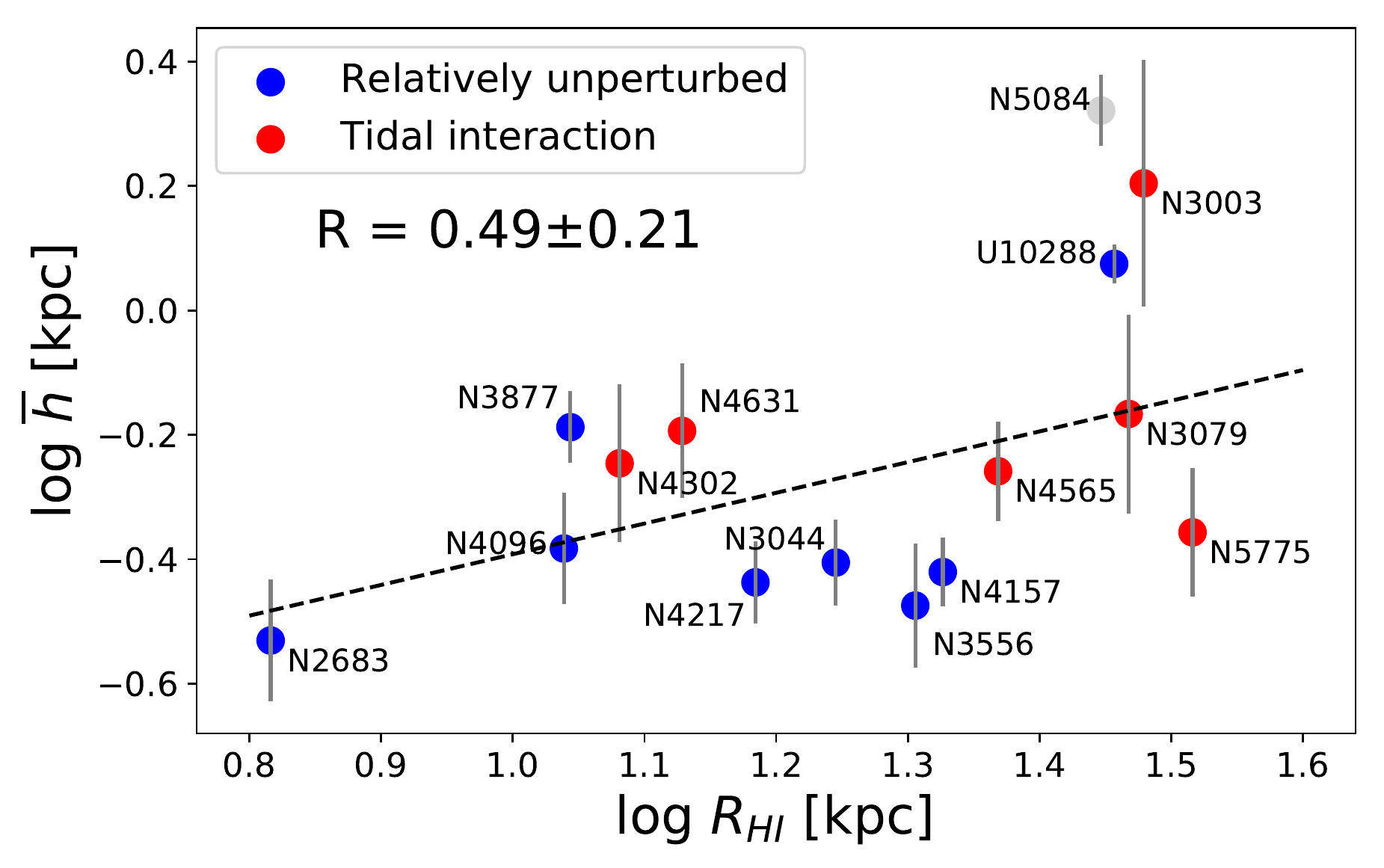}
   \includegraphics[width=0.45\linewidth]{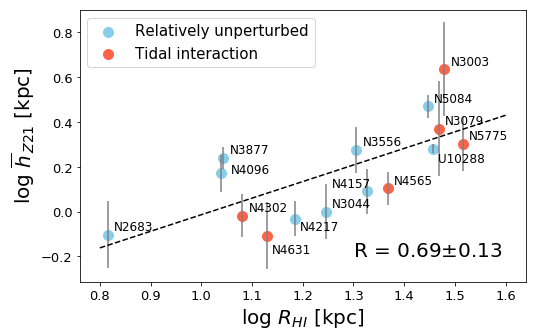}
  \end{minipage}%
 \begin{center}
     {\textbf{(a)}}
 \end{center} 
  \begin{minipage}[t]{0.99\textwidth}
  \centering
   \includegraphics[width=0.45\linewidth]{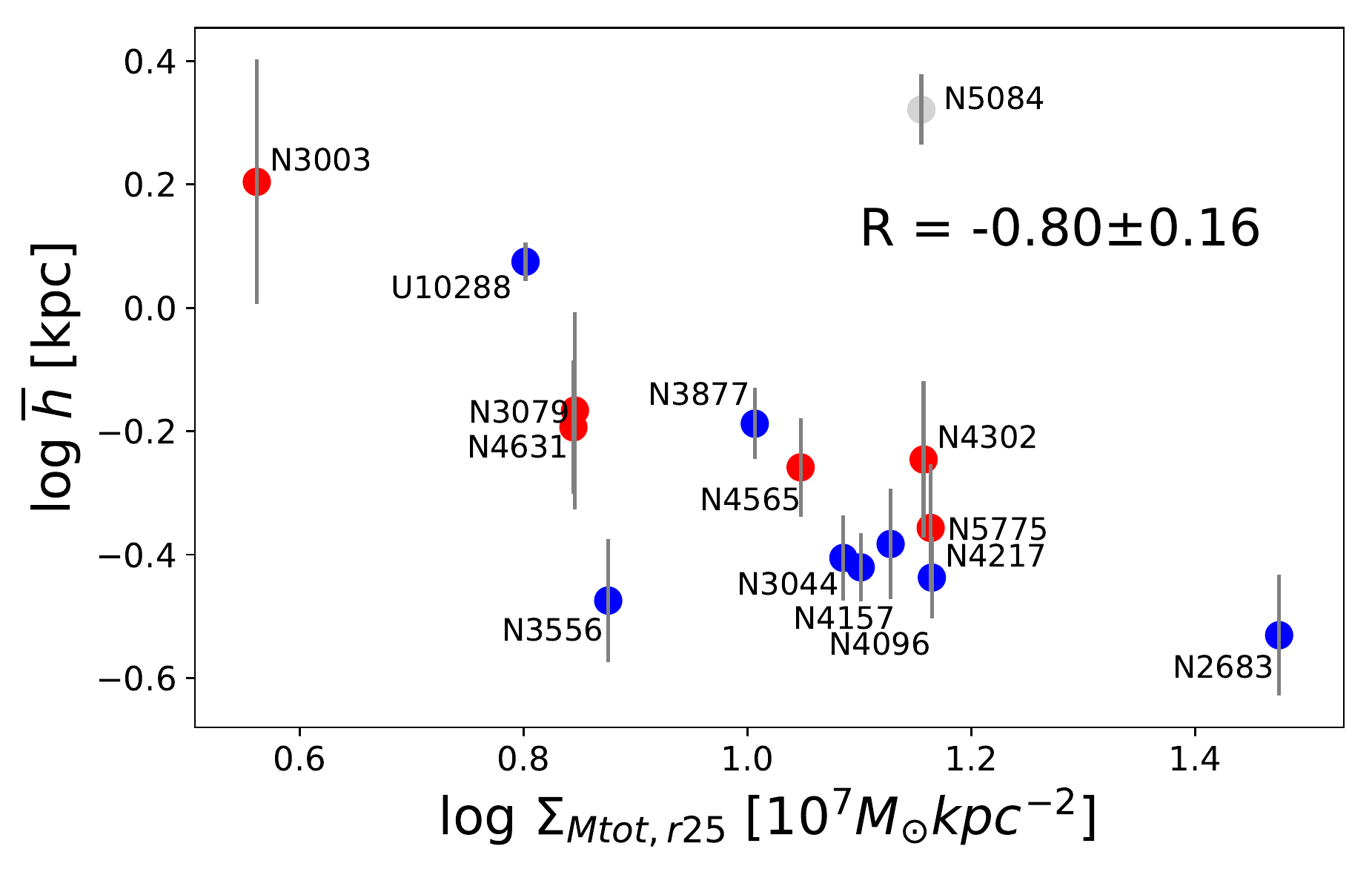}
   \includegraphics[width=0.45\linewidth]{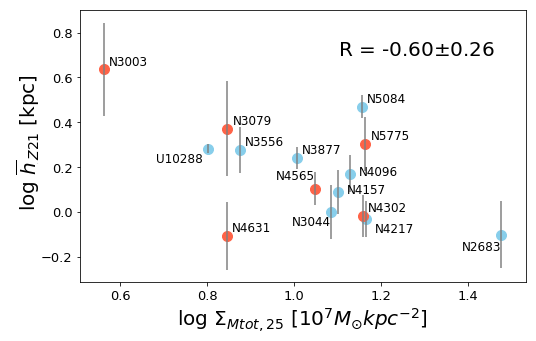}
  \end{minipage}%
  \begin{center}
     {\textbf{(b)}}
 \end{center} 
  \begin{minipage}[t]{0.99\textwidth}
  \centering
   \includegraphics[width=0.45\linewidth]{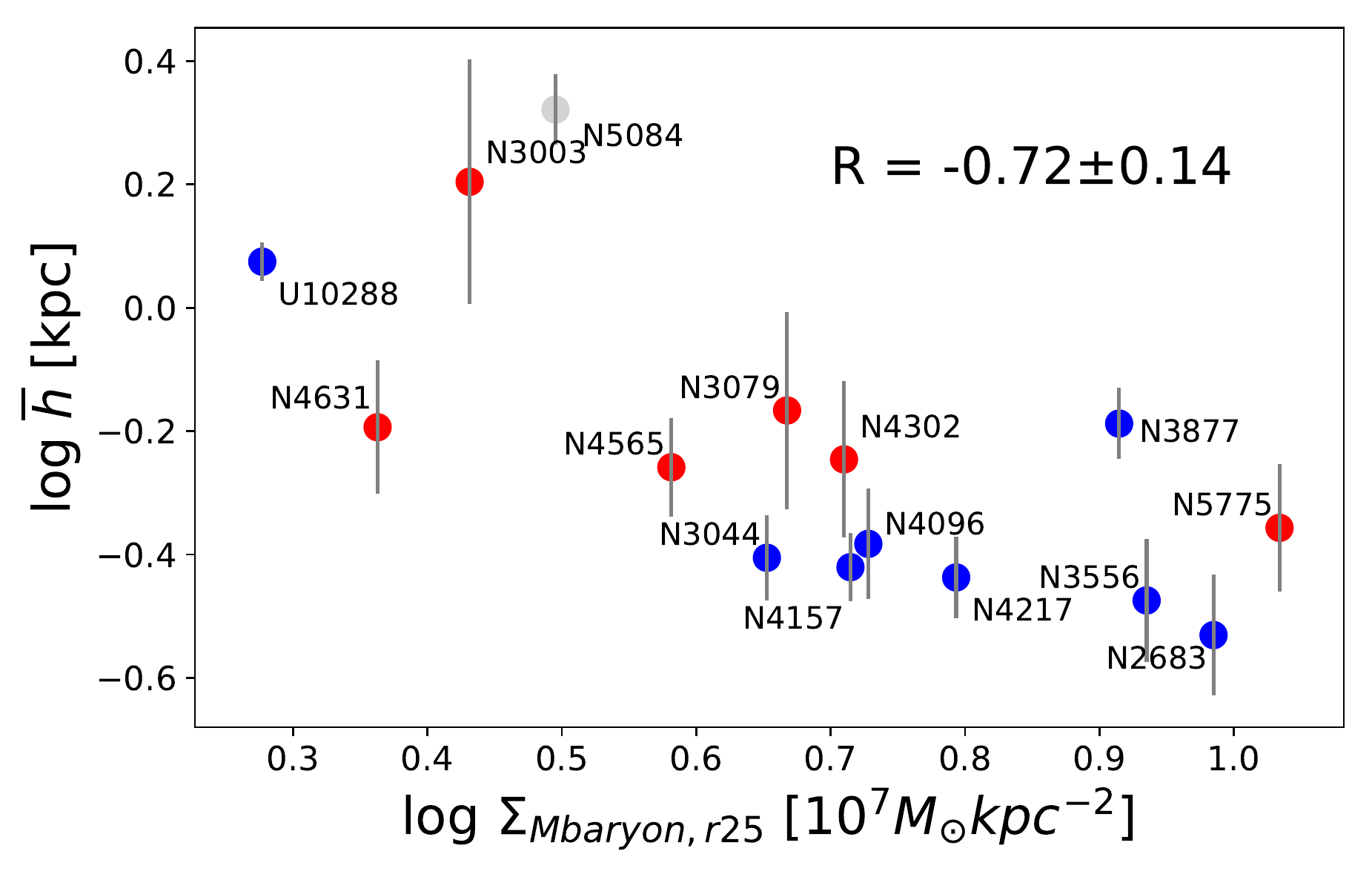}
   \includegraphics[width=0.45\linewidth]{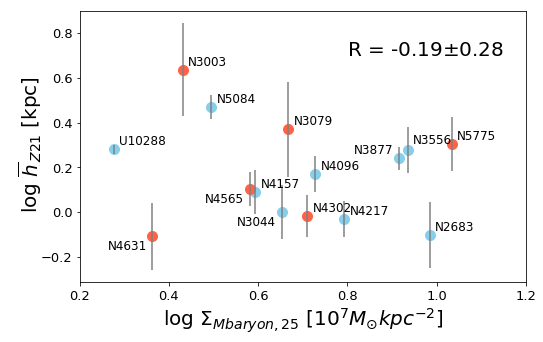}
  \end{minipage}%
  \begin{center}
     {\textbf{(c)}}
 \end{center} 
\caption{The relation between the average $\hi$ Gaussian scale height ($\bar{h}$) and (a) $\hi$ radius ($R_{\rm HI}$), (b) total mass surface density ($\Sigma_{\rm Mtot,r25}$), and (c) baryonic mass surface density ($\Sigma_{\rm Mbaryon,r25}$). The left panels show relations of $\bar{h}$ derived in this work, while the right panels show relations of $\bar{h}_{\rm  Z21}$ from Z21.
The blue points represent the relatively unperturbed galaxies, and the red points are obvious tidally interacting galaxies. The Pearson correlation coefficient (for the edge-on sample excluding NGC 5084) of each relation is list in the corresponding panel. The black dash line in (a) presents the best-fit linear relation (for the edge-on sample excluding NGC 5084) of ${\rm log }\bar{h} = 0.49 \ {\rm log} R_{\rm HI} - 0.89$.}
\label{figure:resultmain}
\end{figure}

\begin{figure}
  \begin{minipage}[t]{0.99\linewidth}
  \centering
   \includegraphics[width=0.45\linewidth]{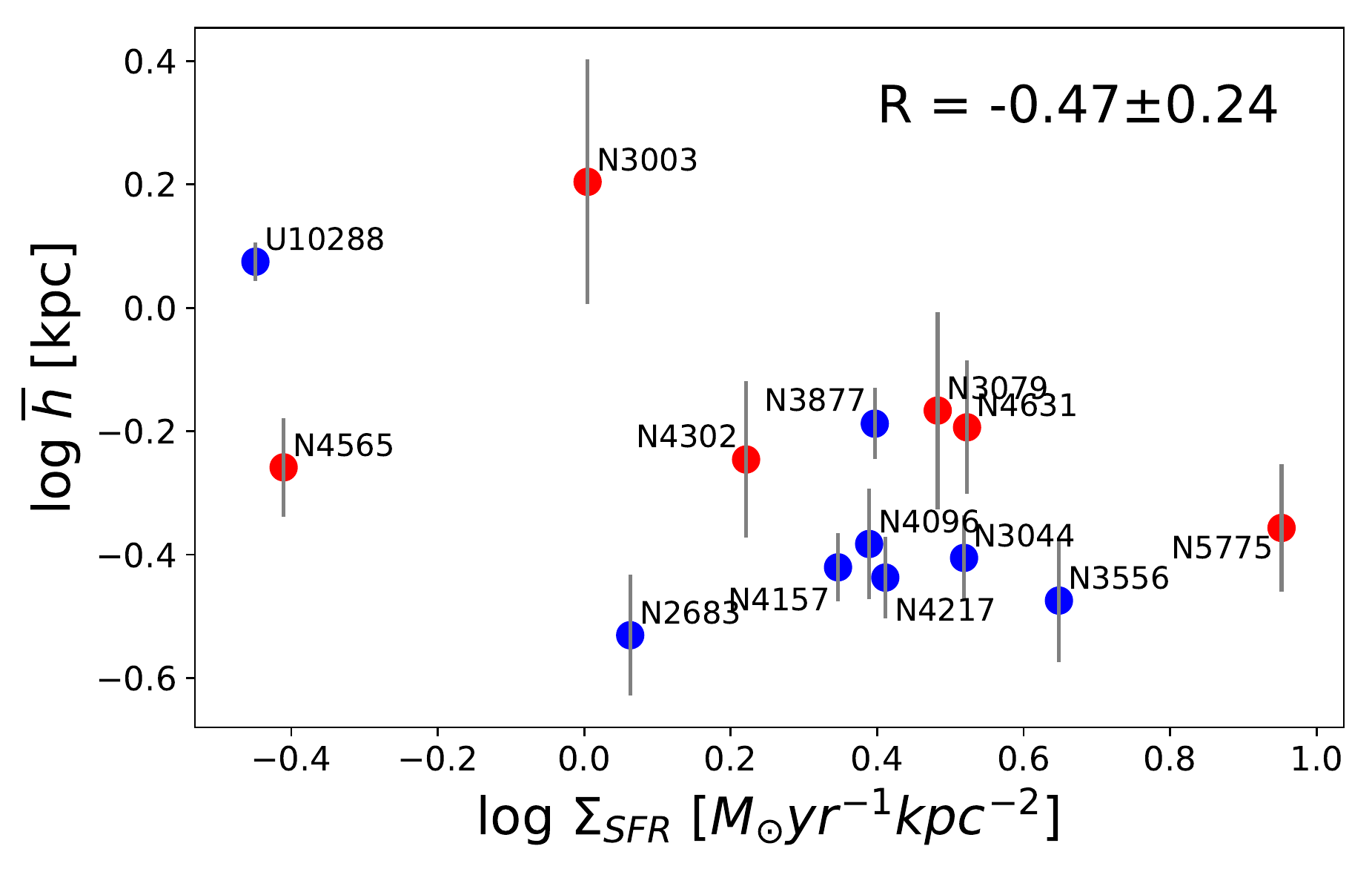}
   \includegraphics[width=0.45\linewidth]{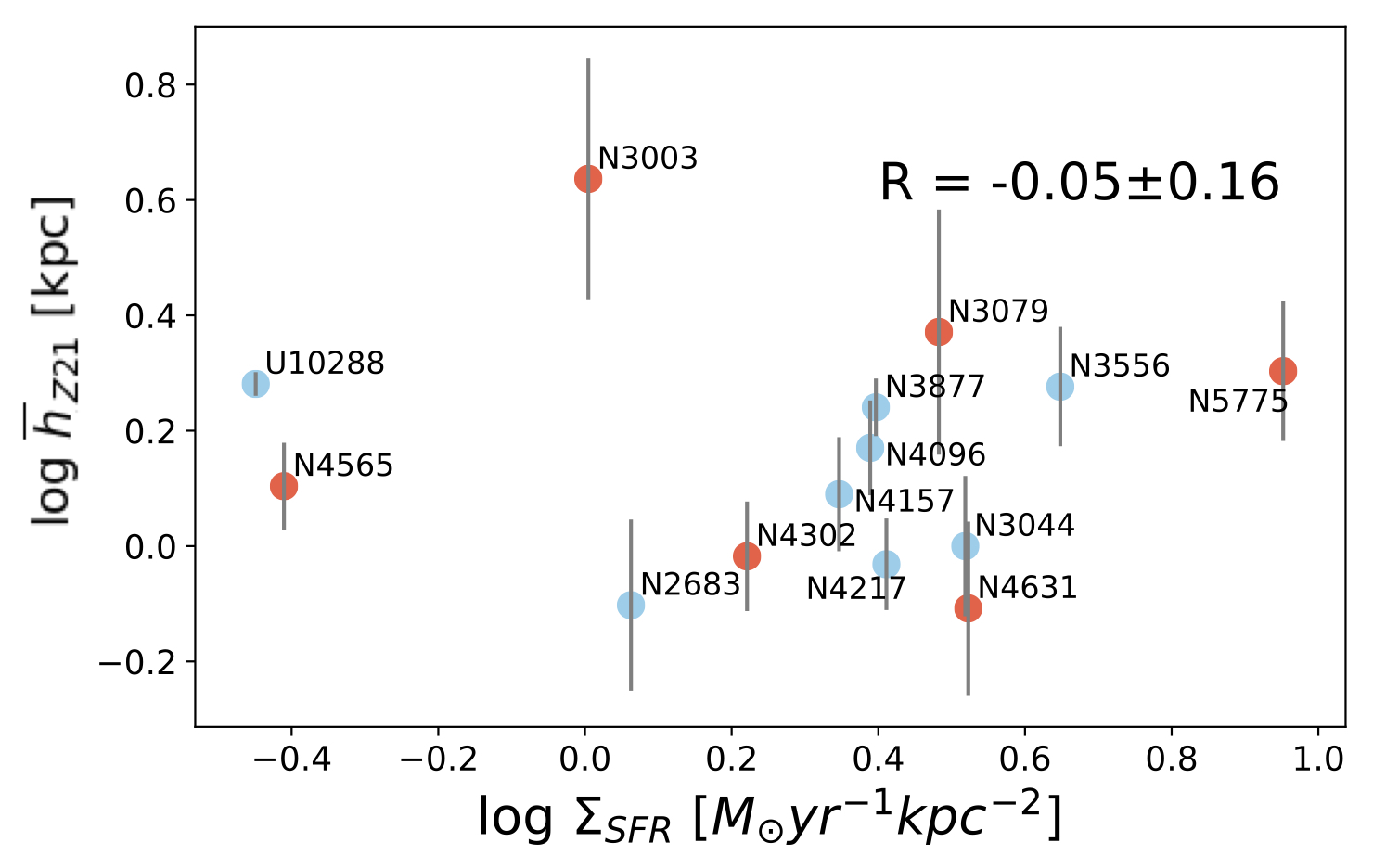}
  \end{minipage}%
  \begin{center}
     {\textbf{(a)}}
 \end{center}
  \begin{minipage}[t]{0.99\textwidth}
  \centering
   \includegraphics[width=0.45\linewidth]{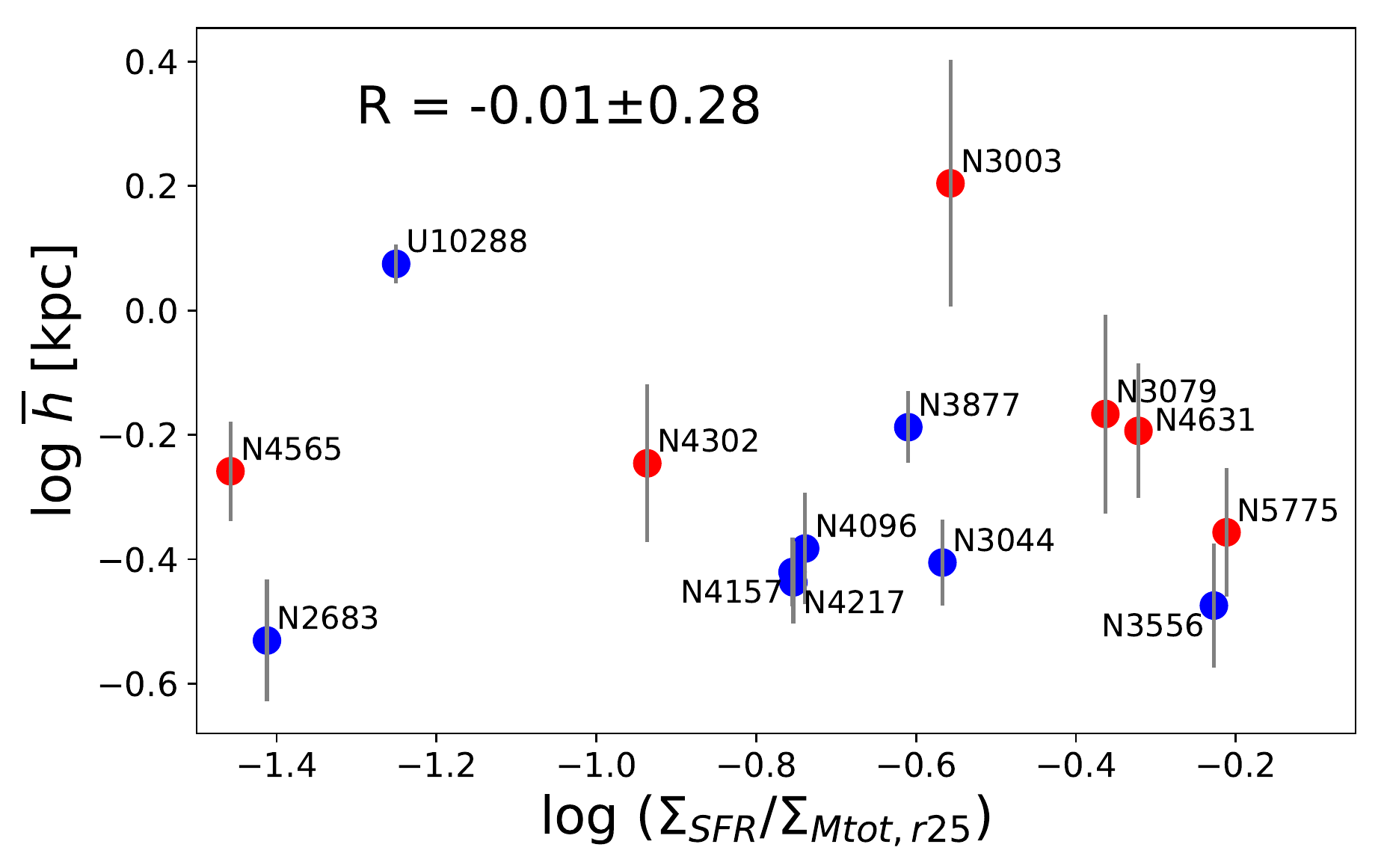}
   \includegraphics[width=0.45\linewidth]{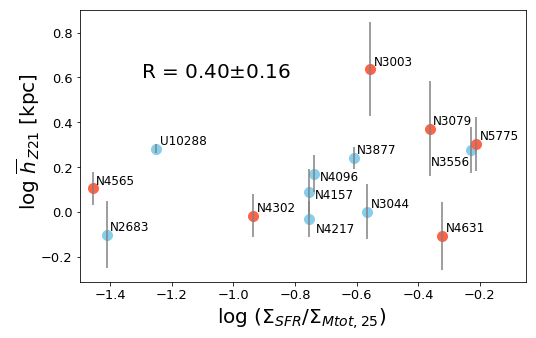}
  \end{minipage}%
  \begin{center}
     {\textbf{(b)}}
 \end{center}
\caption{(a)The relation of the average $\hi$ Gaussian scale height ($\bar{h}$) versus the star formation rate surface density ($\Sigma_{\rm SFR}$). The edge-on sample is separated into high- (pink dots) and low-$\Sigma_{\rm Mtot,r25}$ (dark cyan dots) subsets. (b)The relation of $\bar{h}$ versus the ratio between the star formation rate surface density and total mass surface density ($\Sigma_{\rm SFR}/\Sigma_{\rm Mtot,r25}$). The colors of points represent different environments like in Figure 3. The left panels show the relations of $\bar{h}$ from this work, while the right panels show the relations of $\bar{h}_{\rm  Z21}$ from Z21 like in Figure 3.}
\label{figure:resultsfr}
\end{figure}

We present how the $\hi$ scale heights averaged within the optical radius, $\bar{h}$, correlates with other galactic properties in the left panels of Figures \ref{figure:resultmain} and \ref{figure:resultsfr}. As labeled in the figures, the Pearson correlation coefficient (R) is calculated based on the whole sample excluding NGC 5084.

NGC 5084 is excluded from the correlation coefficient calculation (and fitting for linear relations as well) because of the globally peculiar structure of the whole disk.  Most of its $\hi$ distributes on an outer ring-like structure. This $\hi$ `ring' is following a faint optical outer ring and is tilted with respect to the main optical disk by about 5 deg \citep{1986MNRAS.219..759G, 1990MNRAS.246..324Z}. The origin of the misaligned $\hi$ ring is possibly external gas accretion from satellites, as NGC 5084 is one of the most massive galaxies in its group environment. As the $\hi$ ring is beyond the optical disk, its scale-height is unlikely to be strongly related with the stellar properties. We keep it in the sample, mainly to demonstrate how far galaxies can deviate from major scaling relations when they are globally unsettled. 

For the rest galaxies, we use different colors to separate the tidally interacting systems (red) from the relatively unperturbed galaxies (blue).

The right panels of Figures \ref{figure:resultmain} and \ref{figure:resultsfr} show the corresponding previous results presented in Z21 as comparisons. The averaged Gaussian $\hi$ scale heights from method of Z21 are labeled as $\bar{h}_{\rm  Z21}$ (see Table \ref{table3:hHI}). In Z21, those figures with exponential scale height were presented in Figure 7 and 8 but not discussed in details, for the systematic broadening effects had not been investigated and quantified. In this paper, we can look into the trends in figures in more details.

Figure \ref{figure:resultmain}(a) shows a correlation between $\bar{h}$ and $R_{\rm HI}$. The black dashed line shows the best-fit linear relation ${\rm log }\bar{h} = 0.49 \ {\rm log} R_{\rm HI} - 0.89$. The most significant outliers include two tidal interacting systems (NGC 3003 and NGC 5775), and the peculiar galaxy N5084. From Figure 5 in Z21, NGC 3003 presents signature of recent gas-rich minor merger in the south-west corner. From literature studies, NGC 5775 seems to be in the early stage of a major merger and $\hi$ masses are likely transferred from its neighbor galaxy NGC 5774. This result shows that $\hi$ disks grow thicker when their diameters are larger, which has been found for disks in other wavelengths, for example the optical \citep{1997A&A...320L..21D, 2002AstL...28..527Z} and the radio continuum \citep{2018AA...611A..72K}. In Z21, we obtained a similar relation between $\bar{h}_{\rm  Z21}$ and $R_{\rm HI}$ but with a different slope, for $\bar{h}_{\rm  Z21}$ there were not sufficiently corrected for artificial broadening.

In Figure \ref{figure:resultmain}(b), $\bar{h}$ is significantly anti-correlated with total mass surface density $\Sigma_{\rm Mtot,r25}$. The globally peculiar galaxy NGC 5084 is again an outlier. Interestingly, most of the tidally perturbed galaxies in our sample do not behave as outliers in the relation. It implies that normally tidal interacting effects are not dominating in determining the thickness of those $\hi$ disks. To confirm this speculation, we also test and find no clear correlation between $\bar{h}$ and the local number density of galaxies $\rho$ (taken from Paper I; \cite{2012AJ....144...43I}). This result holds even when we select galaxies with low mass densities. The data points from Z21 show similar strong anti-correlation between $\bar{h}_{\rm  Z21}$ and $\Sigma_{\rm Mtot,r25}$, indicating that the trend is strong enough to show itself above the systematic biases. 

In Figure \ref{figure:resultmain}(c), there is a moderate anti-correlation between $\bar{h}$ and baryonic mass surface density $\Sigma_{\rm Mbaryon,r25}$. The correlation is weaker than that of $\Sigma_{\rm Mtot,r25}$ (implications are discussed in Section 5).
Different from the anti-correlation presented here, the trend of Z21 shows no correlation with R value equal to -0.19 with a scatter of 0.28. This inconsistent result implies that the analysis of the systematic biases in this work is useful and a correction is necessary.

In Figure \ref{figure:resultsfr}(a), for our whole sample, there is no clear correlation between $\bar{h}$ and star-formation rate surface density $\Sigma_{\rm SFR}$. We also study the possible relation of $\bar{h}$ with the ratio between $\Sigma_{\rm SFR}$ and $\Sigma_{\rm Mtot,r25}$ in Figure \ref{figure:resultsfr}(b) and no correlation is observed. This is consistent with Z21 in which no correlation was found between $\bar{h}_{\rm  Z21}$ and $\Sigma_{\rm SFR}$, and between $\bar{h}_{\rm  Z21}$ and $\Sigma_{\rm SFR}/\Sigma_{\rm Mtot,r25}$, when the broadening effects are not as properly corrected for as in this paper.

\section{Discussion on the scaling relations of the scale height}
One of the major results in this paper is that we find the averaged $\hi$ scale height $\bar{h}$ to be significantly anti-correlated with the total mass surface density $\Sigma_{\rm Mtot,r25}$. Such a relation seems largely expected from the vertical hydrostatic equilibrium model of $\hi$, as gravity restores the $\hi$ to the mid-plane of the disk. However, in theory, only the mass within the disk layers contributes effectively to the gravity that restores the $\hi$ \citep{2012ApJ...754...48F, krumholz2018unified}, which our mass density parameters do not strictly indicate. For $\Sigma_{\rm Mtot,r25}$, the dark matter certainly extends beyond the disk layers, though there has been evidence that dark matter halos can be oblate in shape and thus concentrates more mass in the disk layers \citep{1996ApJ...462..563N, 1997ApJ...490..493N}.
On the other hand, as typically observed in other massive galaxies \citep{2000ApJ...533L..99M, catinella2012galex, 2018MNRAS.476.5127R}, the baryonic-to-total mass ratio is relatively high in our edge-on sample ($\sim 0.35 \pm 0.15$), leaving space for uncertainties of the dark matter geometry.
$\Sigma_{\rm Mtot,r25}$ can be viewed as an upper limit of the gravitational mass restoring the $\hi$ disk in the vertical direction. 
The weaker anti-correlation between $\bar{h}$ and baryonic mass surface density $\Sigma_{\rm Mbaryon,r25}$ implies that $\Sigma_{\rm Mbaryon,r25}$ is not a perfect indicator of the restoring gravity when averaging over the radial range of the whole optical disk. 
The total mass is not commonly available for extra-galaxies, and thus as a result it is typically assumed that dark matter is not important and the baryonic mass is sufficient to explain most of the effects related to gravity within the optical radius of massive spiral galaxies. However, our results suggest that the effect of dark matter cannot be ignored at least for $\hi$ thickness.
Typically, stars are dynamically much hotter than the neutral gas \citep{1981A&AS...46..193H}, particularly so in massive galaxies \citep{2010A&A...524A..98W, 2014prpl.conf..149T, 2015ASSL..412...43K}. We may hence overestimate the gravitational contribution from stars, but on the other hand, we have neglected the molecular gas which may have compensated the over-estimated stellar contribution. 
There are also other uncertainties in the relation between $\bar{h}$ and $\Sigma_{\rm Mtot,r25}$ ($\Sigma_{\rm Mbaryon,r25}$). For example, the Poisson equation in cylindrical coordinates indicates that the vertical gravitational acceleration not only depends on the local mass density within a given thickness, but also depends on the changing rate of the disk rotation curve (e.g. Eq. 13 of \cite{krumholz2018unified}). However, our galaxies have high $M_{*}$ and massive galaxies typically have steeply rising and then flattened rotation curves \citep{2008AJ....136.2648D}, minimizing the effective gravitational contribution from this term.
Despite all the caveats discussed above, $\bar{h}$ anti-correlates with $\Sigma_{\rm Mbaryon,r25}$ and particularly strongly with $\Sigma_{\rm Mtot,r25}$, indicating the CHANG-ES $\hi$ disks are strongly regulated by gravity. 

Here we only make a preliminary discussion and hypothesis on the results with $\Sigma_{\rm SFR}$, due to our limited sample size. Stronger conclusions should be made through larger dataset in the future. The supernova feedback and streaming motions of gas are both important channels to produce turbulence \citep{Elmegreen_2003, 2011MNRAS.413..149S, 10.1093/mnras/stw011, 10.1093/mnras/stx2866} and thus raise the $\hi$ disk thickness, and both channels are expected to be associated with star formation. Yet, we do not find a significant relation between $\bar{h}$ and SFR surface densities. A possible reason is that the intrinsic trend is weak and buried under uncertainties, as SFR can be fueled in other ways than gas streaming motions, supernova feedback comes from stars formed earlier than reflected by the current SFR, and a large portion of energy from the supernova feedback may be distributed into gas of other phases. The magnetic pressure may also support the disk vertical structures \citep{krumholz2018unified}; although magnetic properties have been derived for several galaxies in the CHANG-ES program \citep{2019A&A...623A..33S, 2020A&A...639A.112K, 2020A&A...639A.111S}, directly quantifying the magnetic pressure remains to be done.

Our results of $\hi$ scale heights appear to correspond with the study of \cite{2018AA...611A..72K} (K18 hereafter) about the scale heights in the radio continuum. K18 found the radio scale heights correlate with the halo radius (See Figures 12, 13, and 14 in K18), anti-correlate with mass surface densities (See Figures 15, and 16 in K18), and do not correlate with SFR surface densities (See Figure 19 in K18) both in C- and L-band. These consistent results reflect the similar way of organizing their vertical structures for the different constituents of the ISM. In addition, we provide strong support for arguments in K18 that the underlying gravitational potential plays a more important role than star formation in determining disk scale heights. Star formation may be important in determining scale heights and structures in individual locations, but globally, the potential appears to dominate.

\begin{figure}
  \centering
   \includegraphics[width=0.45\linewidth]{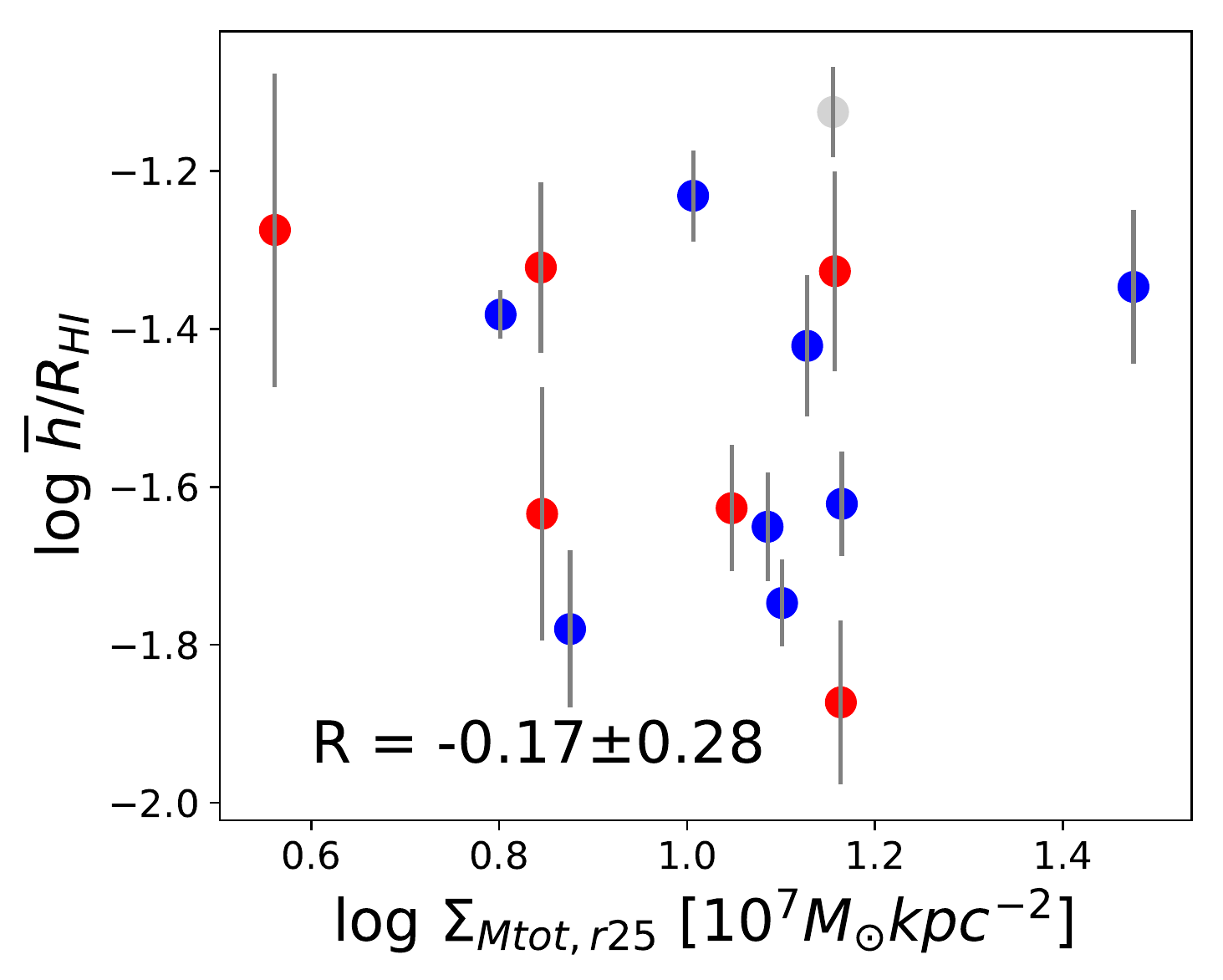}
   \includegraphics[width=0.45\linewidth]{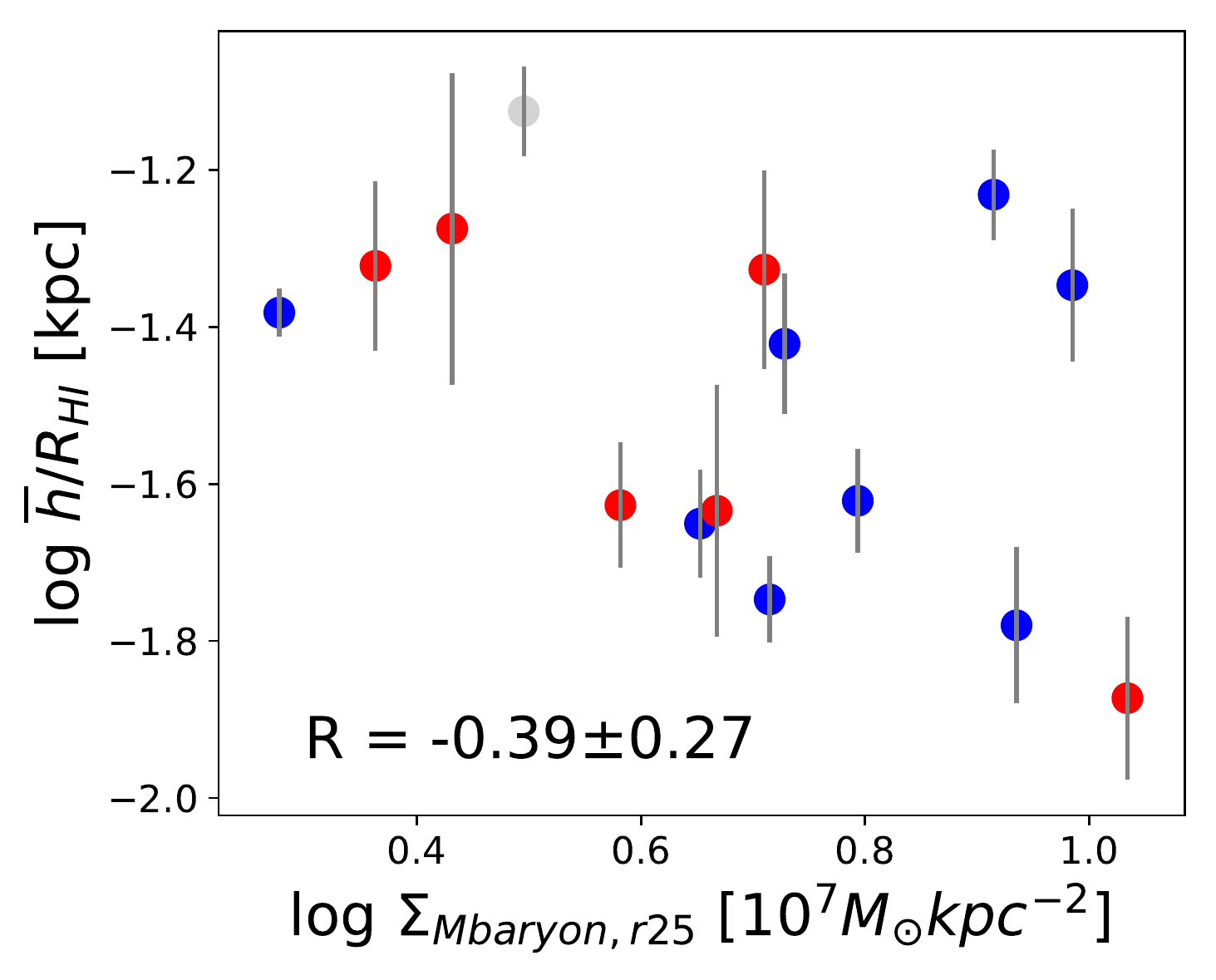}
   \includegraphics[width=0.45\linewidth]{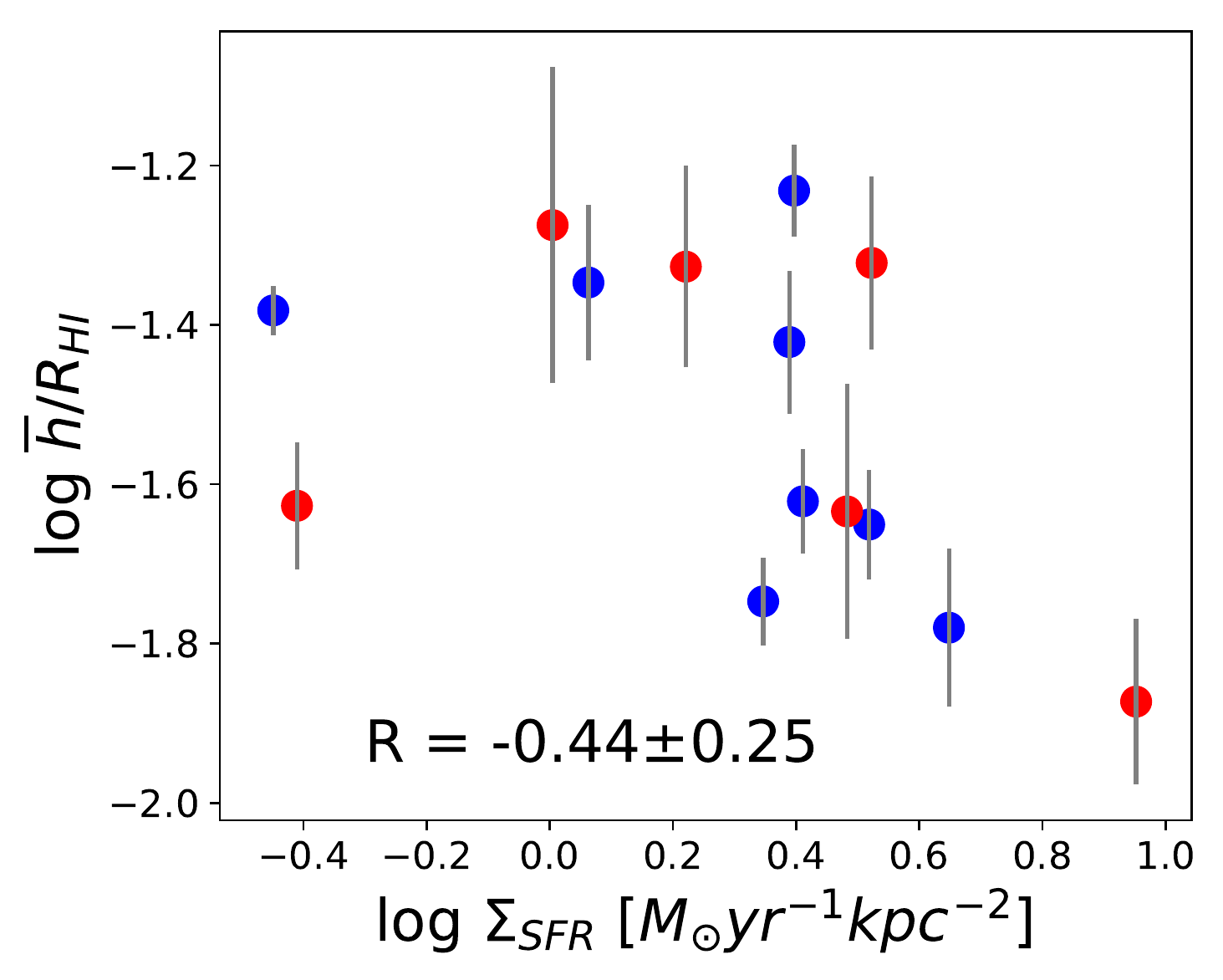}
   \includegraphics[width=0.45\linewidth]{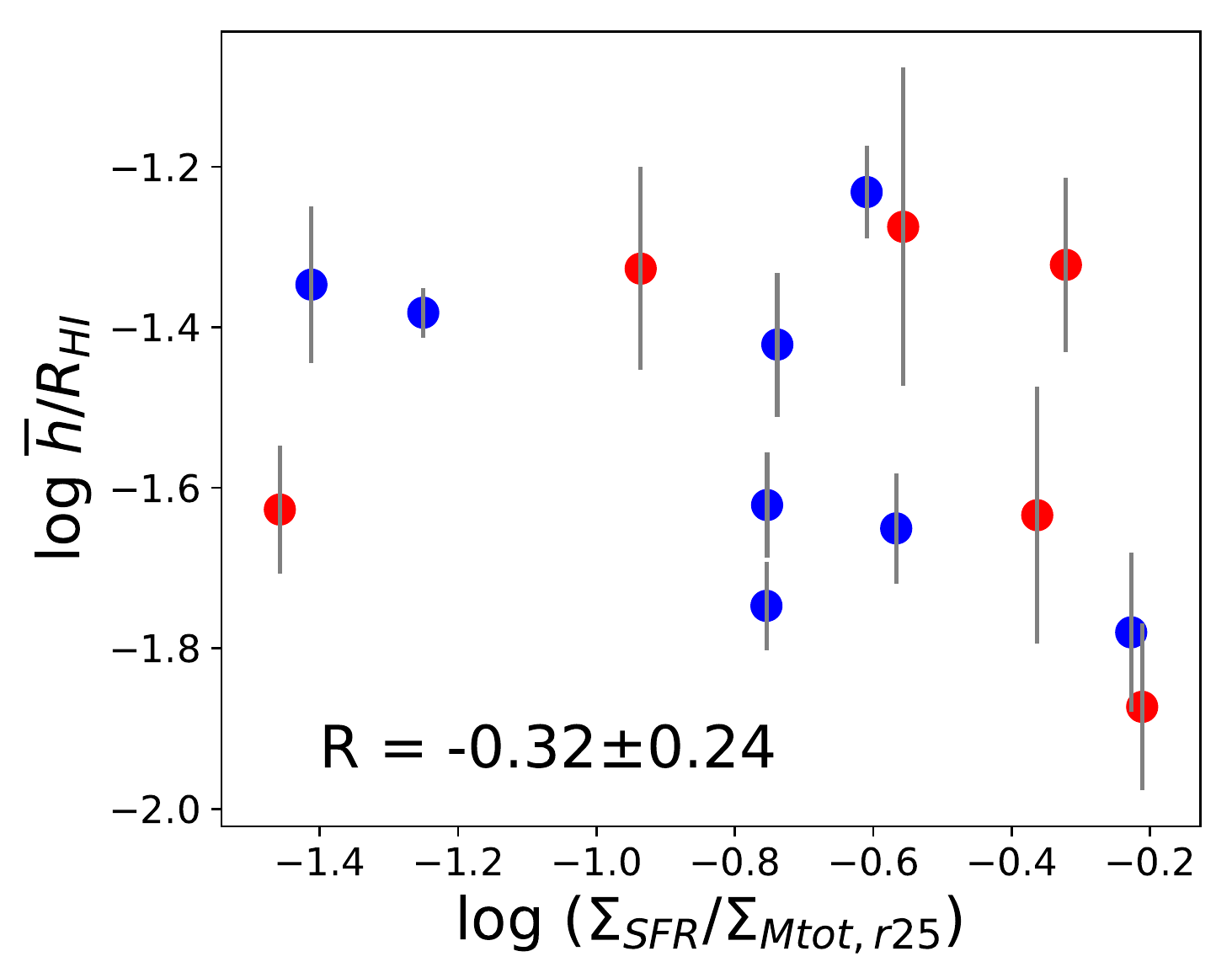}
\caption{The relation of $\bar{h}/R_{\rm HI}$ versus (a) the total mass surface density ($\Sigma_{\rm Mtot,r25}$), (b) the baryonic mass surface density ($\Sigma_{\rm Mbaryon,r25}$), (c) the star formation rate surface density ($\Sigma_{\rm SFR}$), and (d) the ratio between the star formation rate surface density and total mass surface density ($\Sigma_{\rm SFR}/\Sigma_{\rm Mtot,r25}$). The color and label here are the same as Figure \ref{figure:resultmain} and \ref{figure:resultsfr}.}
\label{figure:dis}
\end{figure}

The really strong correlation between $\bar{h}$ and $R_{\rm HI}$ raises the possibility that other dependencies of $\bar{h}$ shown in this study are artificially caused by a mutual dependence on $R_{\rm HI}$. Following the study in Z21, we thus test by showing the correlation between the normalized $\hi$ scale heights $\bar{h}/R_{\rm HI}$ and other galactic properties in Figure \ref{figure:dis}. The $\bar{h}/R_{\rm HI}$ shows no correlation with $\Sigma_{\rm Mtot,r25}$, $\Sigma_{\rm Mbaryon,r25}$, $\Sigma_{\rm SFR}$, and $\Sigma_{\rm SFR}/\Sigma_{\rm Mtot,r25}$. The same results were presented in Z21 and the right column of Figure 8, where the systematic biases are not corrected. These results imply that, it is indeed possible that the relation between $\bar{h}$ and $\Sigma_{\rm Mtot,r25}$ is caused by the more intrinsic correlation between $\bar{h}$ and $R_{\rm HI}$ and are not independent or new. On the other hand, this dependence may have been blurred by other dependencies (e.g., $\bar{h}$ vs. $\Sigma_{\rm SFR}$) which have contrary slopes. More data will be needed to test both hypothesises in the future.

In the end, we re-iterate that the biggest caveat of our analysis is that, due to the lack of kinematic information, we only obtain a rough measure of $\hi$ scale heights averaged over a large radial range, and miss the typical flaring feature of galactic gas disks. The systematic uncertainty due to averaging needs to be quantified in the future, based on edge-on $\hi$ disks with properly modeled 3-d distributions.

\section{Summary and conclusion}

We measure $\hi$ scale height from 15 highly inclined (inclination$> 80^{\circ}$) galaxies from the CHANG-ES sample. We build mock data cubes and images for the $\hi$, to explore the feasibility of deriving disk thickness in the photometric way. We show that when the inclination of disks are not perfectly 90 degree, the directly measured disk scale-heights can be systematically overestimated as a result of planar projection, beam smearing, and disk flaring. We quantify the trends of these over-estimating effects, which are used as correcting equations for the directly measured disk scale heights. We derive the radially averaged Gaussian scale height within the optical radius for each galaxy and apply these corrections. We find a significant anti-correlation of the corrected scale height with the total mass surface density and the baryonic mass surface density. Compared with the results in Z21, these results imply that either the systematic broadening effects are corrected well in our method, or the underlying physical drivers is strong enough to show itself above the biases.

The result is consistent with predictions from the quasi-equilibrium model (Poisson equation) of $\hi$ vertical distribution, where the gravity is the primary force restoring the $\hi$ to the mid-plane of the disk\citep{krumholz2018unified}.


\begin{acknowledgements}
We thank L. Bai, K. Zheng, X. Feng, Y. Fu, especially thank J. Zhu, Y. Yang, N. Yu, Z. Liang, H. Xu. J. Wang acknowledges support from the National Science Foundation of China (12073002, 11721303). Parts of this research were supported by High-performance Computing Platform of Peking University. This research made use of Photutils, an Astropy package for detection and photometry of astronomical sources \citep{larry_bradley_2020_4044744}. {NOD3 \citep{2017A&A...606A..41M}, Photutils \citep{larry_bradley_2020_4044744}, BBarolo \citep{2015MNRAS.451.3021D}}

\end{acknowledgements}

\bibliographystyle{raa}
\bibliography{main}

\end{document}